\documentclass[%
 reprint,
 amsmath,amssymb,
 aps,
]{revtex4-2}

\usepackage{graphicx}% Include figure files
\usepackage{dcolumn}% Align table columns on decimal point
\usepackage[toc,page]{appendix}
\usepackage{bm}% bold math
\usepackage{xcolor}
\DeclareMathOperator{\sech}{sech}

\newcommand{\g}{g_{1,\infty}}

\begin{document}

\preprint{APS/123-QED}

\title{Coupling-independent, Real-time Wireless Resistive Sensing through Nonlinear  $\mathcal{PT}$-symmetry}

\author{Siavash Kananian}
\author{George Alexopoulos}
\author{Ada S. Y. Poon}
\affiliation{Department of Electrical Engineering, Stanford University, Stanford, CA 94305}

\date{\today}% It is always \today, today,

\begin{abstract}
We report the realization of coupling-independent, robust wireless sensing of fully-passive resistive sensors. $\mathcal{PT}$-symmetric operation obviates sweeping, permitting real-time, single-point sensing. Self-oscillation is achieved through a fast-settling nonlinearity whose voltage amplitude is proportional to the sensor's resistance. These advances markedly simplify the reader. A dual time-scale theoretical framework generalizes system analysis to arbitrary operating conditions and a correction strategy reduces errors due to detuning from $\mathcal{PT}$-symmetric conditions by an order of magnitude.

\end{abstract}

\maketitle

%\tableofcontents

\textit{Introduction.---}The discovery that a large subclass of quantum mechanical systems exhibiting non-Hermitian properties possesses entirely real eigenspectra has spurred renewed investigations into coupled-resonator systems \cite{Bender_RealSpectra,ElGanainy_NonHermitian}. Contradicting the Dirac-von Neumann axioms, non-Hermitic systems exhibit purely real eigenspectra provided they are pseudo-Hermitic \cite{Mostafazadeh}, or more specifically, jointly $\mathcal{PT}$-symmetric (invariant to joint spatial reflection and time reversal) \cite{Bender_RealSpectra,Bender_Hamiltonians,Mostafazadeh}. A range of spectral phenomena and applications has recently been observed in coupled electronic resonator systems, including coherent perfect absorption \cite{Sun_Microwave,Schindler_Electronics}, directed transport \cite{Bender_Asymmetric,Fleury_Acoustic}, anti-$\mathcal{PT}$-symmetry \cite{Choi_Anti}, wireless power transfer \cite{Fan_WPT,Fan_WPT_2020}, and the focus of this work: wireless sensing \cite{Ho_EP,Chen_Generalized,Sakhdari_Sensors,Hajizadegan_Inductive,Alu_Glucose}.

\begin{figure}[b]
    \includegraphics[scale = 0.35]{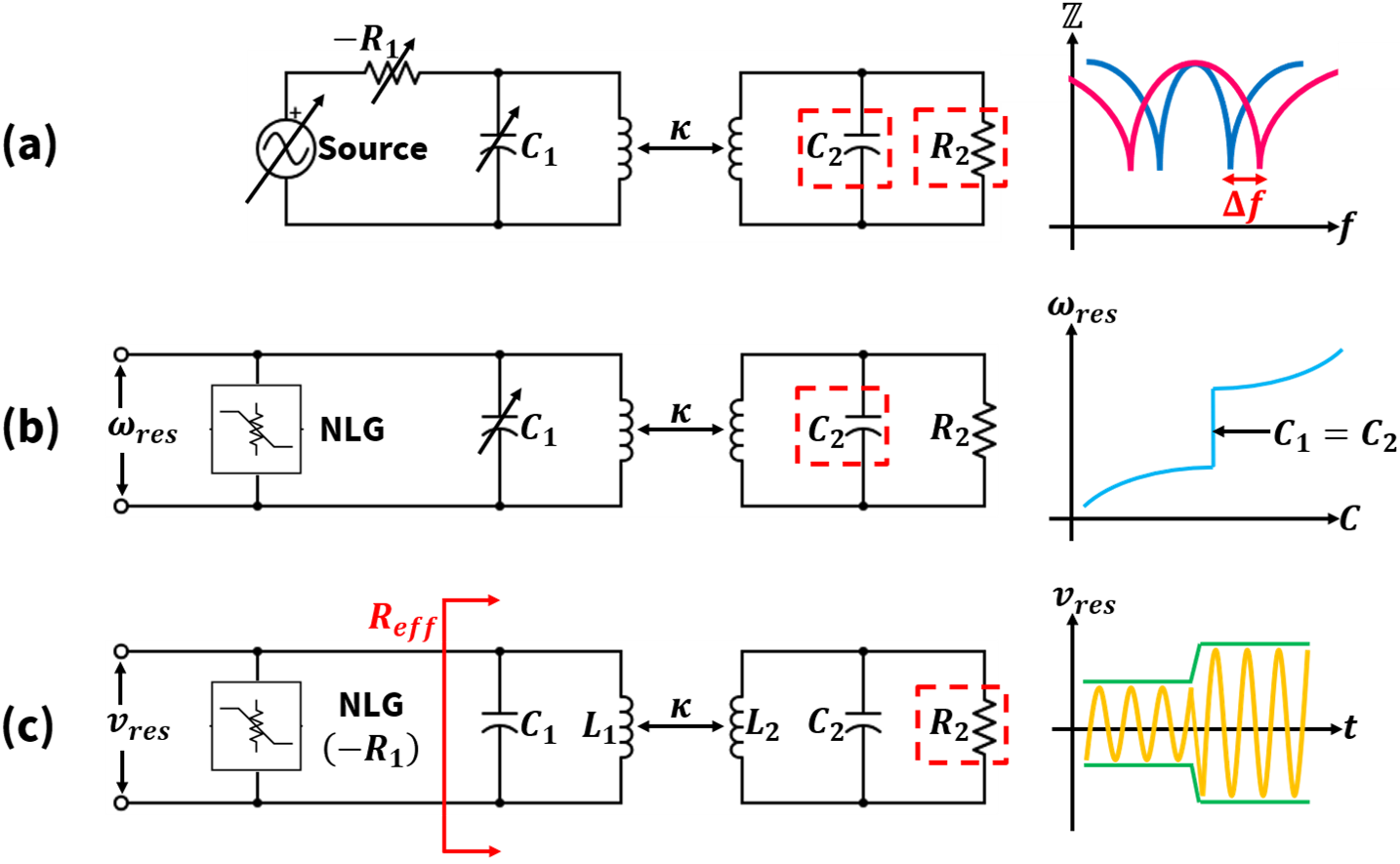}
    \caption{\label{fig:SweepMethod} (a) Forced excitation methods sweep a capacitance, a negative resistance, and the excitation frequency to induce measurable spectral changes in impedance profiles. (b) A nonlinear gain (NLG) element reduces complexity, requiring only one sweep for resonant frequency-based capacitive sensing. (c) The proposed resistive sensing method obviates sweeping through constant operation at exact $\mathcal{PT}$-symmetry; amplitude measurements identify changes in the effective resistance, $R_{eff}$, due to fluctuations in the sensor resistance, $R_2$. Red boxes indicate the sensing element.}
\end{figure}

The dynamics of $\mathcal{PT}$-symmetric operation generate a symmetric pitchfork bifurcation in the eigenfrequency spectrum, contingent on two rigorous conditions: equal resonant frequencies in both resonators (frequency-equalization) and strict gain/loss balance \cite{Bender_RealSpectra,Schindler_Electronics}. Existing $\mathcal{PT}$-symmetric electronic systems adopt a variety of techniques to achieve these conditions, falling under two broad categories: those with forced excitation sources and those with nonlinear self-oscillating gain mechanisms. In the former [Fig.~\ref{fig:SweepMethod}(a)], capacitance sweeps equate resonant frequencies while negative resistance tuning realizes gain/loss balance. Subsequent frequency sweeping of a complex excitation source, typically a network analyzer (VNA), provides sensing through the measurement of spectral fluctuations in the impedance profile \cite{Chen_Generalized,Sakhdari_Sensors,Hajizadegan_Inductive}, which are coupling dependent. Alternatively, a nonlinear gain allows for automatic gain/loss balance and self-oscillation, obviating the need for gain sweeping and forced excitation \cite{Schindler_Electronics, Fan_WPT, Fan_WPT_2020, Ho_EP, Hassan_Nonlinear} [Fig.~\ref{fig:SweepMethod}(b)]. Provided the initial gain induces exponential growth, the system undergoes transient evolution such that the steady-state gain automatically matches the effective loss. A capacitance sweep then provides frequency-equalization, enabling capacitive sensing \cite{Ho_EP}. The reliance on sweeping in both approaches prohibits real-time wireless sensing as each sweep point requires a finite transient settling time; a \emph{single-point sensing} method is therefore desirable as it simplifies readout and achieves real-time operation.

This Letter demonstrates that wireless resistive sensing can be achieved by operation at the point of symmetric bifurcation (exact $\mathcal{PT}$-symmetry) where the effective resistance seen by the gain element is automatically \emph{coupling-independent} and equal to the fully-passive sensor's resistance [Fig.~\ref{fig:SweepMethod}(c)].  The adoption of a nonlinear gain further provides for self-oscillation. As a whole, no sweeping is required, reducing reader complexity and leading to real-time, single-point measurements. A fast-settling nonlinear gain is introduced; steady-state voltage amplitude sensing at this gain element detects the sensor's resistance. In contrast to prior efforts whose exact oscillation amplitude and nonlinearity profile do not affect operation \cite{Schindler_Electronics, Bender_Asymmetric,Fan_WPT, Ho_EP}, our approach dramatically simplifies resistive sensing. We demonstrate that self-oscillation remains even when the system is not exactly $\mathcal{PT}$-symmetric; an error-correction technique is introduced to enhance the robustness of sensing.

\textit{Resistive sensing with coupled resonators.---}Consider the coupled parallel-parallel resonator topology in Fig.~\ref{fig:SweepMethod}(c) with resonant frequencies $\omega_1=1/\sqrt{L_1C_1}$ and $\omega_2=1/\sqrt{L_2C_2}$ where one resonator has gain, $g_1 = R_1^{-1}\sqrt{L_1/C_1}$, and the other has loss, $\gamma_2 = R_2^{-1}\sqrt{L_2/C_2}$. Define the coupling coefficient as $\kappa = M/\sqrt{L_1L_2}$, and the inductance and capacitance ratios as $\mu = \sqrt{L_1/L_2}$ and $\chi= \sqrt{C_1/C_2}$, respectively. Applying Kirchoff's Current Law (KCL), the charges on each resonator, $q_1$ and $q_2$, and their derivatives are related through the coupled equations,
\begin{subequations}\label{eq:1}
\begin{align}
    \ddot{q}_1 - g_1(q_1)\Dot{q}_1 + \frac{1}{1-\kappa^2}q_1 -  \frac{\mu\kappa\chi^2}{1-\kappa^2}q_2 &= 0, \label{subeq:1a} \\
    \ddot{q}_2 + \gamma_2\mu\chi\Dot{q}_2 - \frac{\mu\kappa}{1-\kappa^2}q_1 + \frac{\mu^2\chi^2}{1-\kappa^2}q_2 &= 0,\label{subeq:1b} 
\end{align}
\end{subequations}
where $g_1\bigl(q_1(t)\bigr)$ models the crucial time-varying nonlinear gain. Eqs.~\eqref{eq:1} can be recast into the Liouvillian formalism, $\frac{d}{d\tau}\mathbf{Q} = \mathcal{L}\mathbf{Q}$, where $\mathbf{Q} = \begin{bmatrix}q_1 & q_2 & \Dot{q}_1 & \Dot{q}_2\end{bmatrix}^T$, $\mathcal{L}$ is the Liouvillian matrix of system parameters, and $\tau = \omega_1 t$. This formalism is based on exact circuit-level analysis; hence, the low-$\kappa$ and low-$\gamma_2$ approximations made in couple-mode theory (CMT) that would otherwise restrict the dynamic range and accuracy of wireless resistive sensing, are avoided \cite{Choi_Anti,Fan_WPT,Fan_WPT_2020}.

The coupled system exhibits two time scales~\cite{strogatz_book}: a fast-time governing the steady-state frequency and gain/loss balance of the sinusoidal oscillations corresponding to resistive sensing; and a slow-time, over which the amplitude envelope settles, dictating the sensing speed.

\textit{Fast-time scale.---}Assuming time harmonic solutions, $e^{i\omega_{\lambda} t}$, we find the eigenfrequencies, $\omega_{\lambda}$, using the characteristic equation, $\det(\mathcal{L}-i\omega_{\lambda} \mathbf{I})=0$. The real modes are solved by setting the real and the imaginary parts of the characteristic polynomial to zero,
\begin{subequations}\label{eq:2}
\begin{align}
    &(1-\kappa^2)\omega_{\lambda}^4+\omega_{\lambda}^2 \bigl[\g\gamma_2\rho(1-\kappa^2)-1-\rho^2 \bigr] + \rho^2 = 0, \label{subeq:2a} \\
    &\g = \gamma_2\rho \frac{1-\omega_{\lambda}^2(1-\kappa^2)}{\rho^2-\omega_{\lambda}^2(1-\kappa^2)},\label{subeq:2b} 
\end{align}
\end{subequations}
where $\rho = \mu\chi = \omega_2/\omega_1$. In Eq.~\eqref{subeq:2b}, $\g$ is the steady-state value of the nonlinear gain implemented by the negative resistance; specifically, $\g = \lim_{t\rightarrow\infty} g_1\bigl(q_1(t)\bigr)$. The effective resistance due to the sensor as seen by the negative resistance, $R_{eff}$, is completely cancelled by the steady-state negative resistance; that is, $R_{eff}=\g^{-1} \sqrt{L_1/C_1}$. Self-oscillating modes are obtained by substituting $\g$ from Eq.~\eqref{subeq:2b} into Eq.~\eqref{subeq:2a}. The resulting equation can be reduced to a third-order equation, suggesting one or three real modes depending on the system parameters $\kappa$, $\gamma_2$, and $\rho$. Eqs.~\eqref{eq:2} also reveal that real modes are possible even absent $\mathcal{PT}$-symmetric conditions, provided that the gain automatically adjusts to the value in Eq.~\eqref{subeq:2b}.

\begin{figure}[t]
    \includegraphics[scale = 0.24]{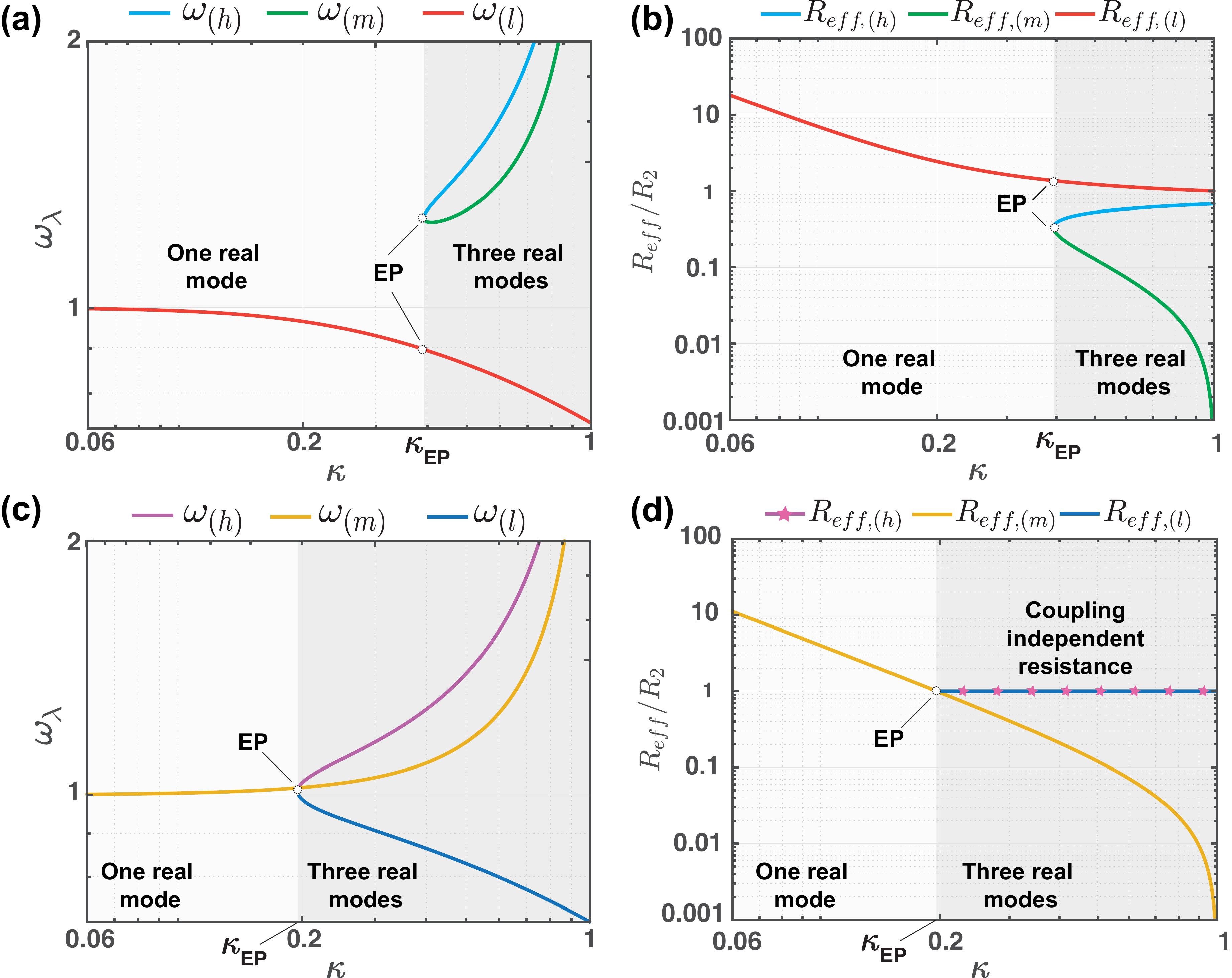}
    \caption{ \label{fig:ParPar_ModesReff} (a) Real modes, $\omega_{\lambda}$, and (b) normalized effective resistance, $R_{eff}/R_2$, vs. $\kappa$ for $\omega_2 = 1.1 \omega_1$ and $\gamma_2=0.2$. (c) Real modes and (d) normalized effective resistances vs. $\kappa$ for $\omega_2 = \omega_1$ and $\gamma_2=0.2$. Note the presence of EPs (marked by circles) denoting the transition from one to three real modes. $\omega_{(h,m,l)}$ and $R_{eff(h,m,l)}$ denote the high, middle, and low eigenfrequencies and effective resistances, respectively.}
\end{figure}

From Eq.~\eqref{subeq:2b}, we find the effective resistance,
\begin{equation}\label{eq:3}
   R_{eff} = \frac{R_2}{\chi^2}\frac{\rho^2-\omega_{\lambda}^2(1-\kappa^2)}{1-\omega_{\lambda}^2(1-\kappa^2)}.
\end{equation}
Fig.~\ref{fig:ParPar_ModesReff}(a) depicts the real modes and their corresponding $R_{eff}$ for $\rho \ne 1$. An exceptional point (EP) exists; below $\kappa_{EP}$, only one real mode exists whereas above $\kappa_{EP}$, three real modes exist. At exact $\mathcal{PT}$-symmetry ($\rho = 1$), the dependence of $R_{eff}$ on the coupling coefficient, $\kappa$, is eliminated. Moreover, if $\mu = \chi = 1$, then $R_{eff} = R_2$. Under these conditions, for $\kappa>\kappa_{EP}$, the following steady-state resonant frequencies, $\omega_{(l,m,h)}$, and steady-state saturated gain values, $g_{(l,m,h)}$, arise from Eq.~\eqref{eq:2} [see Supplemental Material \footnote{See Supplemental Material (Section A.1) for additional information}],
\begin{subequations}\label{eq:4}
\begin{align}
    g_{(l,h)} &= \gamma_2 \label{subeq:4a} \\
    g_{(m)} &= \frac{1}{\gamma_2}\biggl(\frac{1}{1-\kappa^2} - 1\biggr) \label{subeq:4b} \\
    \omega_{(l,h)} &=\pm \sqrt{\frac{2 - \xi\mp\sqrt{\bigl(\xi - 2\bigr)^2 - 4\bigl(1-\kappa^2\bigr)}}{2\bigl(1-\kappa^2\bigr)}} \label{subeq:4c} \\
    \omega_{(m)} &=\pm \sqrt{\frac{1}{1-\kappa^2}} \label{subeq:4d}
\end{align}
\end{subequations}
where $\xi = \gamma_2^2\bigl(1-\kappa^2\bigr)$; for $\kappa\le\kappa_{EP}$, only $g_{(m)}$ and $\omega_{(m)}$ emerge. The location of the EP is derived from Eqs.~\eqref{eq:4},
\begin{equation}\label{eq:5}
    \kappa_{EP} = \sqrt{1 - \frac{2\bigl(\gamma_2^2 + 1\bigr) - 2\sqrt{2\gamma_2^2 + 1}}{\gamma_2^4}},
\end{equation}
where $\kappa_{EP}$ defines the minimum coupling, above which mode-splitting occurs and \emph{coupling-independent} sensing is possible [Fig.~\ref{fig:ParPar_ModesReff}(d)]. Below $\kappa_{EP}$, $\omega_{(l,h)}$ branch out into the complex plane while $\omega_{(m)}$ remains purely real; complex modes cannot sustain steady-state oscillation and are henceforth ignored.

At the exact phase ($\kappa>\kappa_{EP}$), the two modes, $\omega_{(l,h)}$, exhibit lower saturated gains, $g_{(l,h)}<g_{(m)}$, and satisfy conservation of energy, whereas $\omega_{(m)}$ does not and is hence unstable [see Supplemental Material \footnote{See Supplemental Material (Section A.2) for additional information}]. Stable oscillation occurs at either of $\omega_{(l,h)}$ with an effective resistance, $R_{eff,(l,h)} = R_2$, independent of $\kappa$. Unlike capacitive sensing, where variations in $C_2$ alter $\omega_2$ (presenting frequency-imbalance), variations in $R_2$ do not affect $\omega_2$. Therefore, $\omega_1$ and $\omega_2$ can be equalized and fixed \emph{a priori}, precluding the need for a time-intensive frequency sweep to detect the condition $\omega_1=\omega_2$ and hence, enabling single-point sensing. Furthermore, prior frequency-swept methods directly measure spectral variations in the modes \cite{Chen_Generalized,Sakhdari_Sensors,Hajizadegan_Inductive}; these are \emph{coupling-dependent}, since $\omega_{(l)}-\omega_{(h)}$ varies with $\kappa$ [Fig.~\ref{fig:ParPar_ModesReff}(c)].

\begin{figure}[t]
    \includegraphics[scale = 0.24]{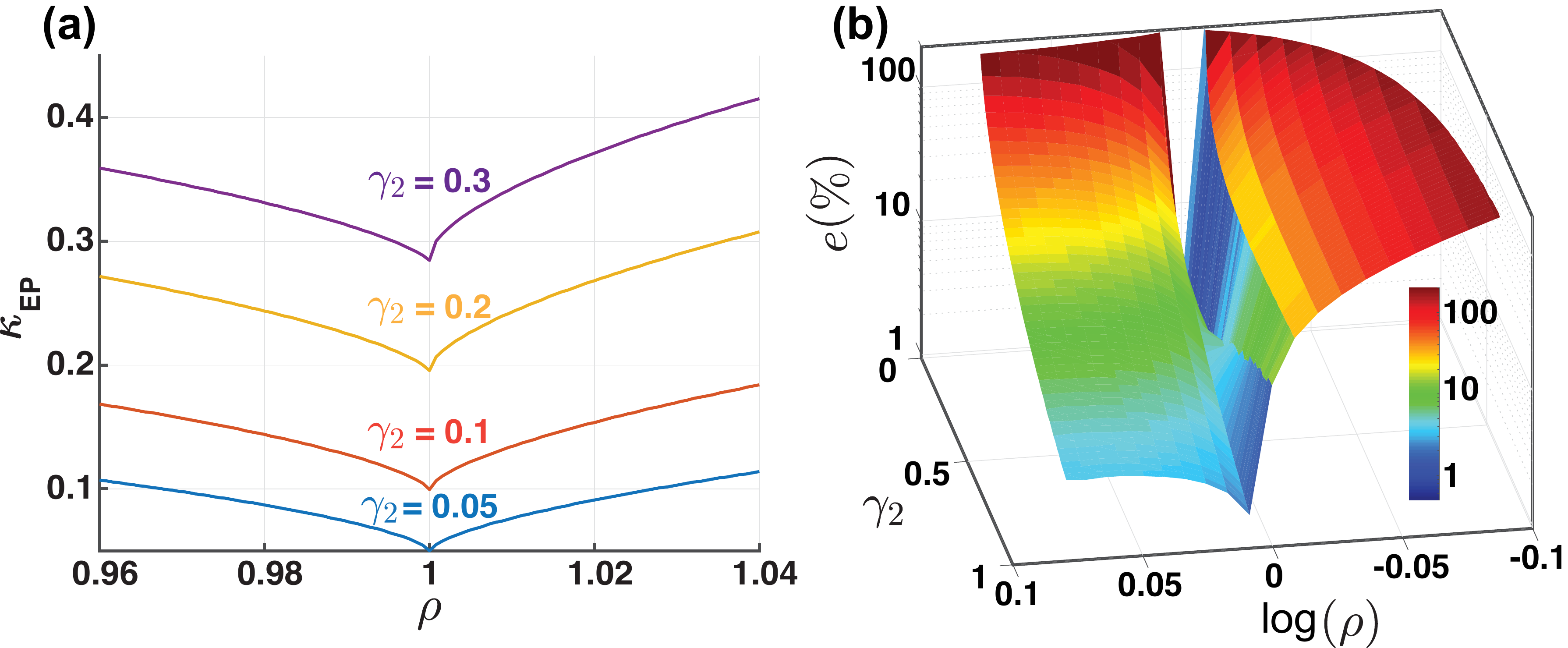}
    \caption{\label{fig:robustness} (a) $\kappa_{EP}$ vs. $\rho$ for four values of $\gamma_2$; note that if $\rho = 1$,  then $\kappa_{EP}\le \gamma_2$, and when $\rho\ne1$, then $\kappa_{EP} >\gamma_2$. (b) Measurement error defined as $e (\%) = |(R_{eff,\kappa=\kappa_{EP}}-R_{2})/R_2|\times 100$ as a function of $\rho$ and $\gamma_2$. For $\rho = 1$, $e=0$; as $\rho$ deviates from unity, $e$ increases.}   
\end{figure}

While here we have focused the discussion on the parallel-parallel resonator topology, we have shown that the series-series resonator topology also achieves coupling-independent sensing beyond the EP [see Supplemental Material \footnote{See Supplemental Material (Section A.3) for additional information}]. However, based on Eq.~\eqref{eq:5}, the desire for a large coupling-independent sensing range restricts $\kappa_{EP}$ and hence $\gamma_2$ ($R_2$). For the series-series resonator topology, $R_2$ is confined to $[0,R_{2,max})$, whereas for the parallel-parallel resonator topology, $R_2$ is confined to a larger range, $(R_{2,min},\infty)$, making it more favorable [see Supplemental Material \footnote{See Supplemental Material (Section A.4) for additional information}]. Finally, series-parallel and parallel-series resonator topologies such as that in \cite{Fan_WPT_2020} are not considered due to their \emph{coupling-dependence} beyond low-$\kappa$ approximations [see Supplemental Material \footnote{See Supplemental Material (Section A.5) for additional information}]. 

\textit{Robust Operation.---}Coupling-independent operation requires identical resonant frequencies in both resonators; this occurs when $\rho = 1$ in Eq.~\eqref{eq:3}. However, naturally-occurring deviations from these conditions induce coupling dependence in $R_{eff}$. Although self-oscillation still arises provided the initial gain is larger than $\gamma_2$, a larger $\kappa_{EP}$ is required [Fig.~\ref{fig:robustness}(a)]. Additionally, maintaining low $\kappa$-dependence requires a larger coupling, limiting the readout range. To capture this deviation, Fig.~\ref{fig:robustness}(b) shows the measurement percent error, $e$, as a function of $\rho$ and $\gamma_2$. For example, for $\rho = 1.15$ and $\gamma_2=0.42$, $e=15\%$, and for $\gamma_2=0.25$, $e=31\%$.

To maintain sensing accuracy, we propose a technique where multiple discrete measurements are taken to mitigate the error due to coupling dependence. Unknown system parameters from~Eqs.~\eqref{eq:2} can be determined through multiple measurements; for example, assuming $\mu = 1$ and known $L_1$, measurements of the mode ($\omega_{\lambda}$) and the coupling-dependent $R_{eff}$ leave $\chi$, $\gamma_2$, and $\kappa$ unknown. By performing measurements at two different coupling strengths, $\kappa_1$ and $\kappa_2$, we can solve a system of four equations and four unknowns ($\mu$, $\gamma_2$, $\kappa_1$, and $\kappa_2$). Additional discrete measurements and post-processing provide enhanced accuracy [see Supplemental Material \footnote{See Supplemental Material (Section B) for additional information}].

\begin{figure}[t]
    \includegraphics[scale = 0.24]{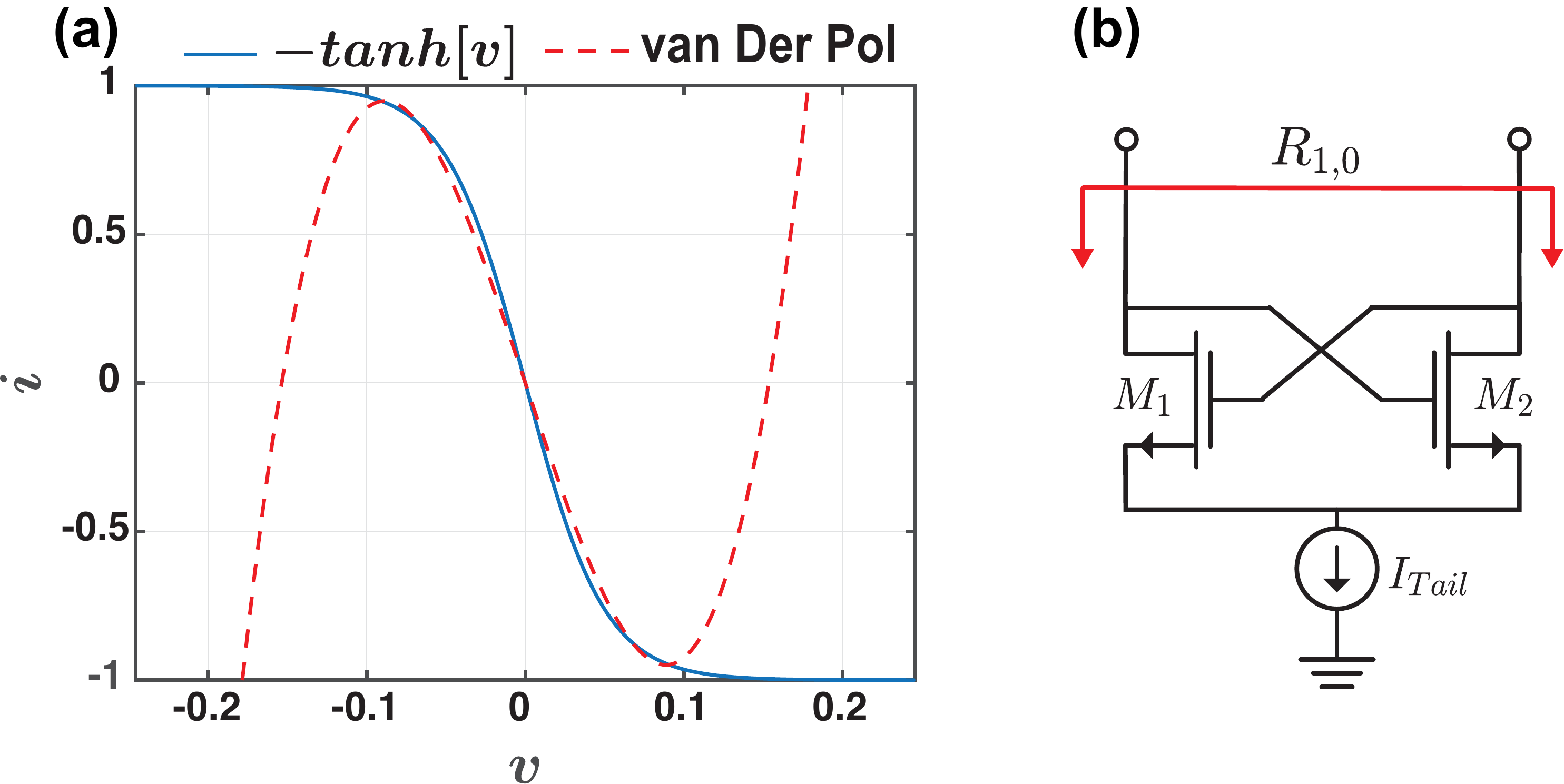}
    \caption{\label{fig:nonlinear} (a) Normalized $i$-$v$ curves for $\tanh$ and van der Pol nonlinearities. (b) The proposed negative resistance circuit with cross-coupled MOSFET pair. }
\end{figure}

\textit{Slow-time scale.---}The transient envelope of the response to Eqs.~\eqref{eq:1} affects the settling time and determines the sensing speed. Understanding this time scale requires a proper model for the nonlinear gain, $g_1(\cdot)$. Traditional models constrain themselves to lower-order van der Pol nonlinearities; however, as a relaxation oscillator, variations in the van der Pol damping term primarily affect the transient waveform shape and the slope of the nonlinearity is not monotonically negative [Fig.~\ref{fig:nonlinear}(a)] \cite{strogatz_book}. Instead, sensing measurements are simplified by monitoring the steady-state voltage amplitude; hence, a monotonically compressive nonlinearity is desirable.

Such nonlinearity can be implemented through the MOS transistor cross-coupled pair circuit \cite{razavi_book} [Fig.~\ref{fig:nonlinear}(b)], whose amplitude, in contrast to previous compressive gain mechanisms \cite{Schindler_Electronics,Fan_WPT}, is not fixed. The MOS cross-coupled pair exhibits a differential current approximated by $2V_T(R_{1,0})^{-1}\tanh{\bigl[(q_1C_1^{-1})/(2V_T)\bigr]}$ where $V_T$ is the thermal voltage and $R_{1,0}$ is the initial negative resistance defined by the transconductance of identical transistors $M_1$ and $M_2$. The charge-derivative of this current gives the dynamic nonlinear model for $g_1(\cdot)$ [see Supplemental Material \footnote{See Supplemental Material (Section C) for additional information}],
\begin{equation}\label{eq:6}
    g_1(q_1) = g_{1,0}\sech^2\bigg[\frac{q_1}{2C_1V_T}\bigg]
\end{equation}
where $g_{1,0} = -(R_{1,0})^{-1}\sqrt{L/C}$ is the initial gain. The transistors switch on and off producing a square-wave that is filtered at the steady-state, resonant frequency \cite{razavi_book_rfic}. From Fourier analysis, the amplitude of the fundamental component of the resulting voltage is $V_1 = (2/\pi)I_{Tail}R_{eff}$ where $I_{Tail}$ is the bias current that sets the initial gain \cite{razavi_book_rfic}. For $\kappa>\kappa_{EP}$, $R_{eff}=R_2$, predicting a \emph{coupling-independent} steady-state amplitude,
\begin{equation}\label{eq:7}
    V_1 = \frac{2}{\pi}I_{Tail}R_2,
\end{equation}
that is directly proportional to $R_2$.

\begin{figure}[t!]
    \includegraphics[scale = 0.24]{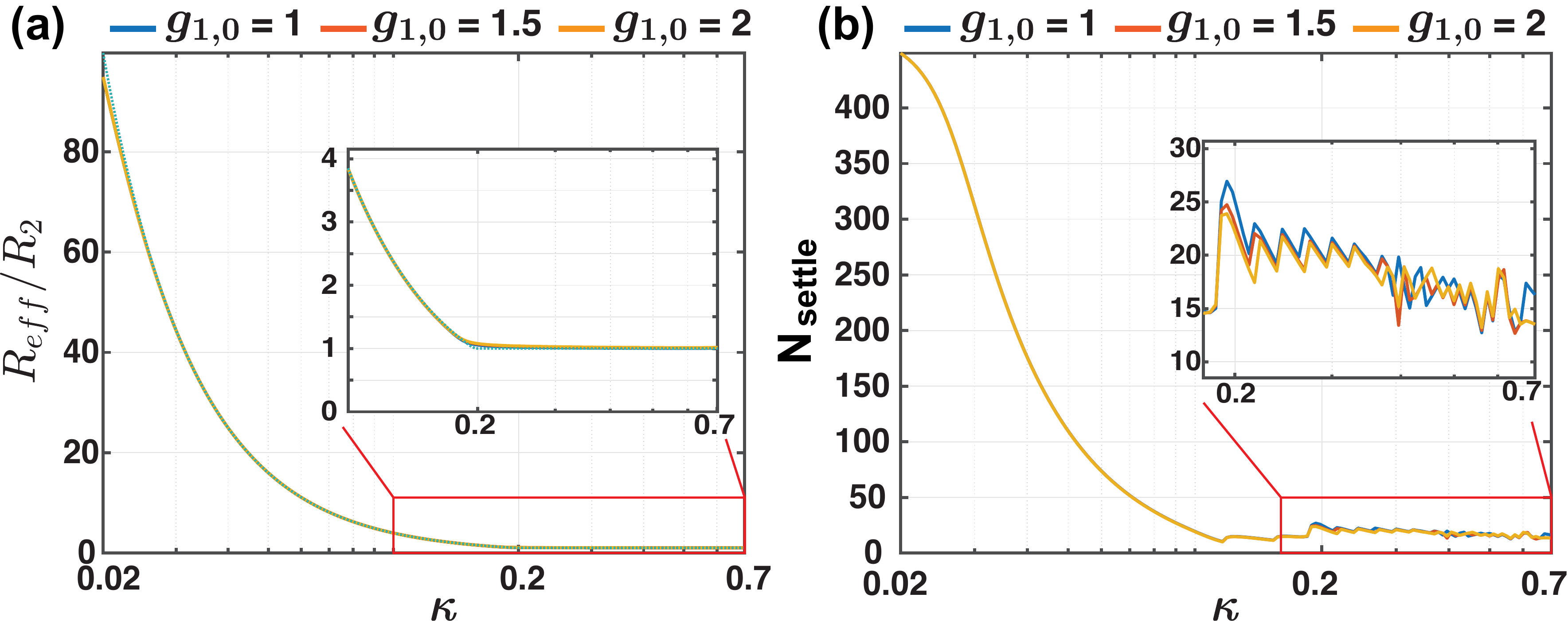}
    \caption{\label{fig:transient} Transient simulations of (a) normalized $R_{eff}$ and (b) $99\%$ settling time in cycles versus coupling for several $g_{1,0}$, and $\gamma_2=0.2$ and $\rho = 1$. The normalized effective resistance is coupling-independent at unity and fast settling is confirmed $\forall\kappa\ge\kappa_{EP}$.}
\end{figure}

Fig.~\ref{fig:transient}(a) shows transient simulations of Eqs.~\eqref{eq:1} with $g_1(\cdot)$ modeled by Eq.~\eqref{eq:6}. The settled steady-state amplitude in Eq.~\eqref{eq:7} demonstrates the coupling-independence of $R_{eff}=R_2$ beyond the EP. The settling time is estimated by measuring the number of cycles it takes for the amplitude to settle within $\pm\delta$ of $V_1$ where $\delta$ represents the desired fraction of settling. With the given nonlinearity, at the exact phase of $\mathcal{PT}$-symmetry, settling times of $25$ cycles suffice for $\delta = 0.01$ [Fig.~\ref{fig:transient}(b)]. For reader resonant frequencies in the High Frequency (HF) range ($>5$ MHz), this corresponds to settling times of $<5\text{ }\mu$s, enabling real-time sensing.

\textit{Experimental Verification.---}The proposed single-point sensing with compressive nonlinearity and self-oscillation allows for a simple reader implementation. As a proof-of-concept, a prototype of the system is built using off-the-shelf components, where the core of the reader circuitry consists of the MOS cross-coupled pair with a programmable capacitor and an inductor implemented using copper traces on a flexible circuit board [Fig.~S10]. The amplitude and frequency of the oscillations are measured using a micro-controller. On the sensor side, an identical inductor and fixed capacitor are used along with a programmable resistor to vary $R_2$, emulating a resistive sensor [Fig.~S11]. The distance between the sensor and the reader is varied over a range of 1~mm to 3~cm.

\begin{figure*}[t!]
 %   \centering
    \includegraphics[scale = 0.28]{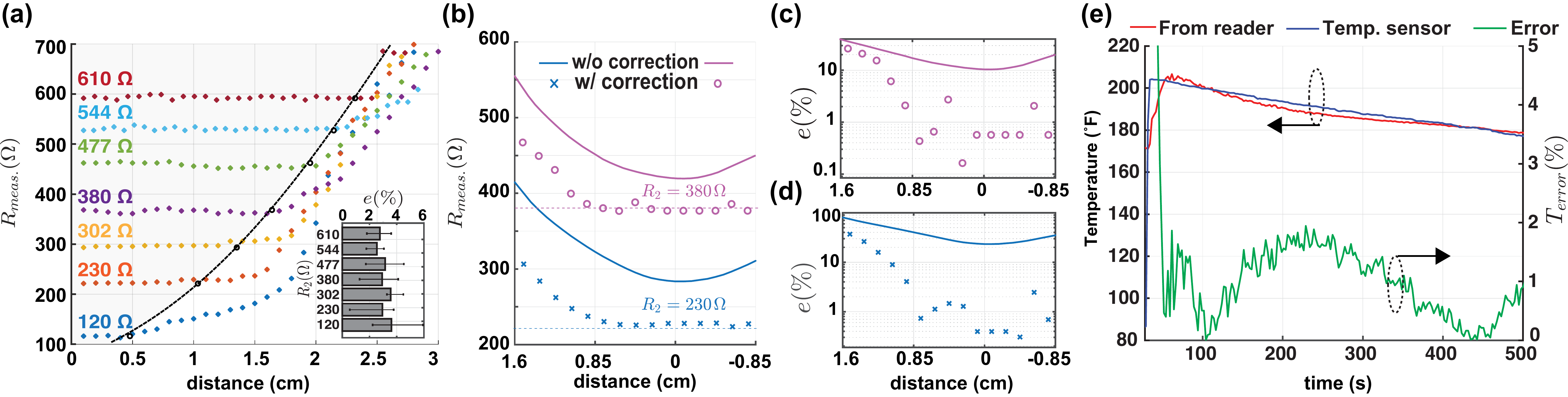}
    \caption{\label{fig:measurement_results} (a) Single-point, real-time measurement results, where the shaded area marks the theoretical coupling-independent sensing region, $\kappa>\kappa_{EP}$, for each $R_2$ setting and the circles show the theoretical distance for $\kappa = \kappa_{EP}$. (b) Multiple-point measurement with imbalanced resonant frequencies ($\rho = 1.15$) while the sensor is moved toward (positive distance) and away (negative distance) from the reader for two $R_2$ settings (230~$\Omega$ and 380~$\Omega$). Measurement error with and without correction for (c) $R_2 = $230~$\Omega$ and (d) $R_2 =$ 380~$\Omega$; error correction reduces the measurement error from $20\%$ to $<1\%$ over a range of 1~cm for $R_2 = $230~$\Omega$ and from $10\%$ to $<1\%$ over a range of 1.5~cm for $R_2 =$ 380~$\Omega$. (e) Real-time measurement of a fully-passive sensor in which a thermistor is employed using the proposed technique along with the measurement from an independent sensor. The percent error of the temperature measurement is shown.}
\end{figure*}

Fig.~\ref{fig:measurement_results}(a) shows measurement results along with the error at each measurement point for each resistance setting. For each setting, the theoretical $\kappa_{EP}$ is calculated using Eq.~\eqref{eq:5} and then converted to distance based on full-wave EM simulations. Next, we replace the fixed sensor capacitor with a variable capacitor to introduce frequency mismatch. In this mode of operation, the reader makes multiple discrete measurements of $(V_{1},\omega_{\lambda})$ at different distances as it moves towards or away from the sensor. A system of four equations and four unknowns is solved for each two consecutive measurements [see Supplemental Material \footnote{See Supplemental Material (Section D) for additional information}]. Fig.~\ref{fig:measurement_results}(b) shows the measurement result for a significant frequency mismatch of $\omega_2 = 1.15 \omega_1$ for two different $R_2$ values. The measurement error over a distance of 1.6~cm is shown in Figs.~\ref{fig:measurement_results}(c)--(d) for the two sensor values. The correction algorithm improves the measurement error by more than an order of magnitude.

\textit{Demonstration of Wireless Sensing.---}The flexible reader is embedded on a paper sleeve to provide real-time wireless measurements of the temperature of hot beverages in a paper cup using a thermistor as a resistive sensor [Fig.~S12]. The sensor exhibits a $4\%$ drop in resonant frequency due to dielectric loading from immersion in water; a scaling factor accounts for this error in measurement. Fig.~\ref{fig:measurement_results}(e) demonstrates wireless sensor measurements, showing that the converted temperature from the sensor faithfully follows that of an independent temperature sensor in real-time.

\textit{Conclusions.---}In this Letter, we show that $\mathcal{PT}$-symmetric operation of a system of two coupled resonators allows for coupling-independent, real-time wireless resistive sensing. We introduce a monotonically compressive nonlinearity in the negative resistance using MOS transistors whose steady-state voltage amplitude tracks the sensor resistance. These techniques obviate the need for parameter sweeps, enabling a low-complexity reader with real-time sensing capability.

The system is analyzed in two time scales: a fast-time governing the modes and gain/loss balance; and a slow-time during which the amplitude envelope settles. Our theoretical framework generalizes system analyses to arbitrary coupling and loss conditions, boosting the sensing dynamic range and accuracy. Additionally, we show that although self-oscillation persists even absent $\mathcal{PT}$-symmetric conditions, error is introduced from the resulting coupling dependence. A correction algorithm based on our fast-time analysis reduces this measurement error by an order of magnitude. A hardware prototype validates our theoretical findings and demonstrates wireless single-point measurement of a fully-passive resistive sensor. Our theoretical framework, nonlinear method, correction algorithm, and simple reader/sensor implementation will ultimately offer an alternative to available technologies such as radio-frequency identification (RFID) and near-field communication (NFC), simplifying the measurement of fully-passive sensors.

\textit{Acknowledgements.---}S. K. and G. A. acknowledge support from the Qualcomm Innovation Fellowship, and A. P. acknowledges support from Chan Zuckerberg Biohub. This work is supported in part by the Stanford Bio-X IIP seed grant.

\appendix

\bibliography{apssamp}% Produces the bibliography via BibTeX.

%%%%%%%%%% Merge with supplemental materials %%%%%%%%%%
\clearpage
\widetext
\begin{center}
\textbf{\large Coupling-independent, Real-time Wireless Resistive Sensing through Nonlinear $\mathcal{PT}$-symmetry}
\vspace{1mm}
\linebreak
\textit{\large Supplementary Material}
\vspace{1mm}
\linebreak
\text{Siavash Kananian, George Alexopoulos, Ada Poon}
\end{center}
%%%%%%%%%% Merge with supplemental materials %%%%%%%%%%
%%%%%%%%%% Prefix a "S" to all equations, figures, tables and reset the counter %%%%%%%%%%

\setcounter{equation}{0}
\setcounter{figure}{0}
\setcounter{table}{0}
\setcounter{section}{0}
\makeatletter
\renewcommand{\theequation}{S\arabic{equation}}
\renewcommand{\thefigure}{S\arabic{figure}}
\renewcommand{\arraystretch}{1.25}
\newcolumntype{P}[1]{>{\centering\arraybackslash}p{#1}}
\newcolumntype{M}[1]{>{\centering\arraybackslash}m{#1}}
\newcommand{\overbar}[1]{\mkern 1.5mu\overline{\mkern-1.5mu#1\mkern-1.5mu}\mkern 1.5mu}
\renewcommand{\thesection}{\Alph{section}}
\renewcommand{\thesubsection}{\Alph{section}.\arabic{subsection}}
\renewcommand{\thefigure}{S\arabic{figure}}
\renewcommand{\theequation}{S\thesection.\arabic{equation}}
%\counterwithin*{equation}{section}

%\begin{appendix}

\section{Fast-time Scale Derivations}
\label{sec:Mode_Derivations}

\begin{figure}[h!]
    \centering
    \includegraphics[scale = 0.46]{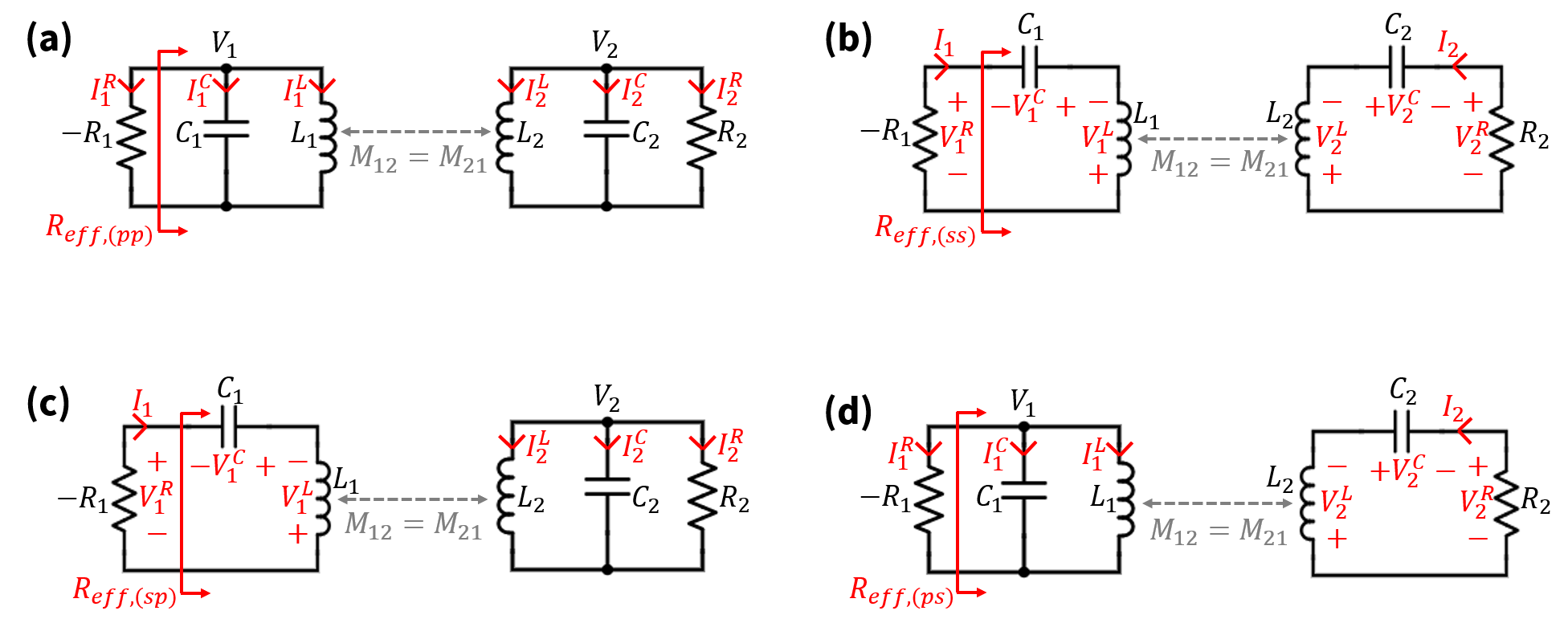}
    \caption{(a) Parallel-parallel, (b) series-series, (c) series-parallel, and (d) parallel-series resonator topologies showing relevant branch currents and loop voltages.}
    \label{fig:AppA_Circuits}
\end{figure}

\noindent Here, we derive the real modes (eigenfrequencies) and effective resistances seen by the negative resistance for four resonator combinations shown in Fig. \ref{fig:AppA_Circuits}. These results are confirmed by circuit impedance analyses.

\subsection{Parallel-Parallel Resonator Topology Derivations}
\label{sec:ModeDer_PP}

\noindent We first consider the parallel-parallel topology [Fig.~\ref{fig:AppA_Circuits}(a)]. We begin by writing capacitor currents as $i_{C_n} = \frac{dq_n}{dt}$ and resistor currents as $i_{R_1} = -v_1/R_1 = -g_1^p\omega_1q_1$ and $i_{R_2} = \gamma_2^p\omega_2q_2$. KCL yields
\begin{subequations} \label{eq:A1_1}
\begin{align}
    i_{L_1} &= g_1^p\omega_1q_1 - \frac{dq_1}{dt}, \\
    i_{L_2} &= -\gamma_2^p\omega_2q_2 - \frac{dq_2}{dt}
\end{align}
\end{subequations}
and the $i$-$v$ relationships at the inductors are given by
\begin{subequations} \label{eq:A1_2}
\begin{align} 
    v_1 = \frac{q_1}{C_1} = L_1\frac{di_{L_1}}{dt} + M\frac{di_{L_2}}{dt} \quad&\Rightarrow\quad \omega_1^2q_1 = \frac{di_{L_1}}{dt} + \frac{\kappa}{\mu}\frac{di_{L_2}}{dt}, \\
    v_2 = \frac{q_2}{C_2} = L_2\frac{di_{L_2}}{dt} + M\frac{di_{L_2}}{dt} \quad&\Rightarrow\quad \omega_2^2q_2 = \frac{di_{L_2}}{dt} + \kappa\mu\frac{di_{L_1}}{dt}.
\end{align}
\end{subequations}
We normalize time by $\tau = \omega_1t$, and substitute Eqs.~\eqref{eq:A1_1} into Eqs.~\eqref{eq:A1_2} to obtain Eqs.~\eqref{eq:1},
\begin{subequations}\label{eq:A1_3}
\begin{align}
    \frac{d^2q_1}{d\tau^2} &= -\frac{1}{1-\kappa^2}q_1 + \frac{\kappa\mu\chi^2}{1-\kappa^2}q_2 + g_1^p\frac{dq_1}{d\tau}, \\
    \frac{d^2q_2}{d\tau^2} &= \frac{\kappa\mu}{1-\kappa^2}q_1 - \frac{\mu^2\chi^2}{1-\kappa^2}q_2 - \gamma_2^p\mu\chi\frac{dq_2}{d\tau}.
\end{align}
\end{subequations}
These equations can be re-written using the Liouvillian formalism:
\begin{equation}\label{eq:A1_4}
    \frac{d}{d\tau}\begin{bmatrix}
    q_1 \\ q_2 \\ \Dot{q_1} \\ \Dot{q_2}
    \end{bmatrix} = \begin{bmatrix}
        0 & 0 & 1 & 0 \\
        0 & 0 & 0 & 1 \\
        -\frac{1}{1-\kappa^2} & \frac{\kappa\mu\chi^2}{1-\kappa^2} & g_1^p & 0\\
        \frac{\kappa\mu}{1-\kappa^2} & -\frac{\mu^2\chi^2}{1-\kappa^2} & 0 & -\gamma_2^p\mu\chi\\
      \end{bmatrix}\begin{bmatrix}
    q_1 \\ q_2 \\ \Dot{q_1} \\ \Dot{q_2}
    \end{bmatrix} = \mathcal{L}
    \begin{bmatrix}
    q_1 \\ q_2 \\ \Dot{q_1} \\ \Dot{q_2}
    \end{bmatrix}.
\end{equation}
\begin{figure}[b!]
    \centering
    \includegraphics[scale = 0.6]{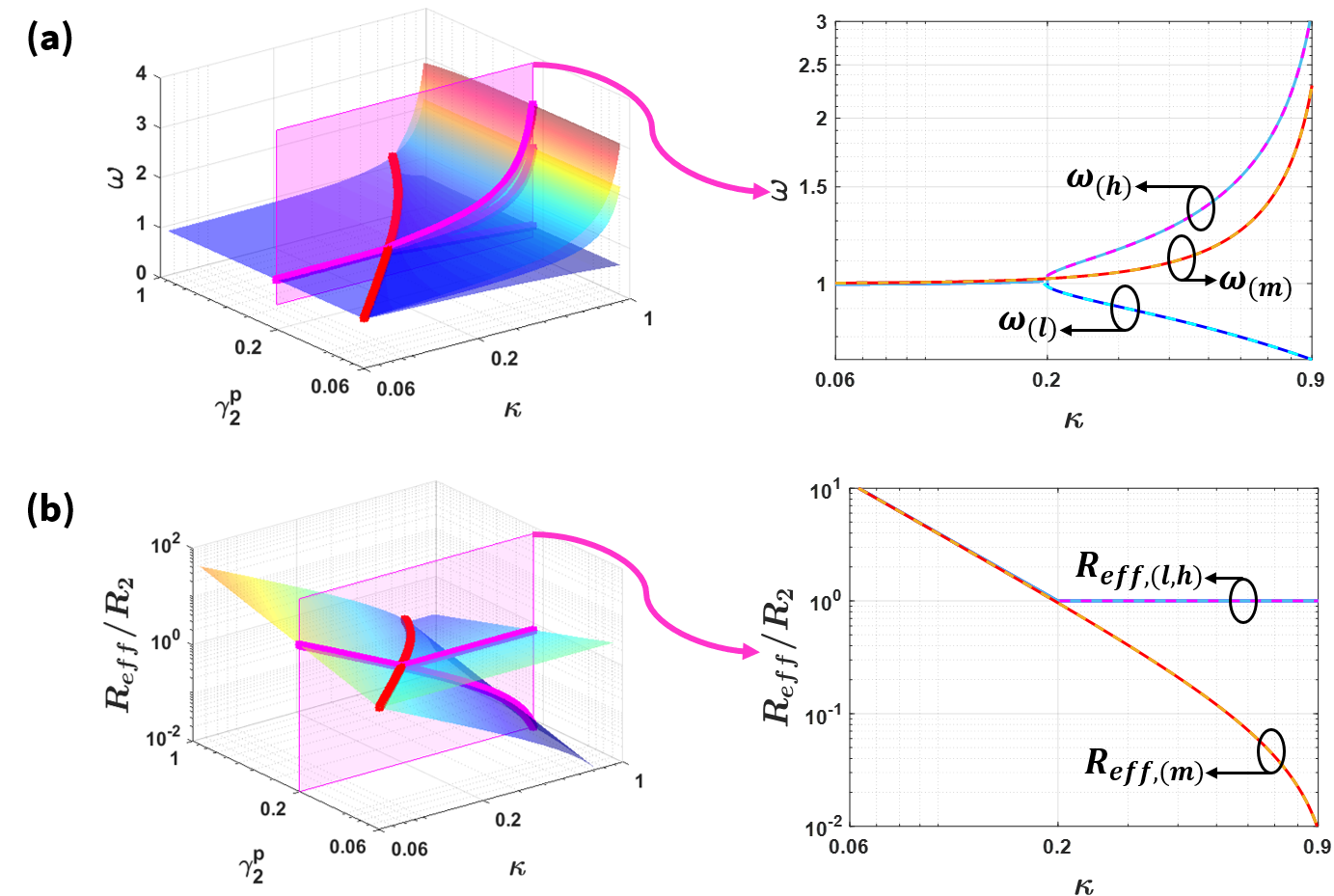}
    \caption{(a) Positive modes and (b) normalized effective resistances corresponding to each mode for the parallel-parallel resonator topology in Fig.~\ref{fig:AppA_Circuits}(a) and $\rho=1$. Slices show solutions for $\gamma_2^p=0.2$. Mode-splitting past $\kappa_{EP}$ (red markers) allows the lower and the upper modes to exhibit a coupling-independent effective resistance. Dashed lines (impedance solutions) confirm Liouvillian solutions (solid lines). The corresponding circuit parameters are $R_2 = 511.25$~$\Omega$, $L = 2.3$~$\mu$H, and $C = 220$~pF.}
    \label{fig:figappA_ppModes}
\end{figure}

\noindent The real and imaginary parts of the characteristic equation for this Liouvillian matrix are,
\begin{subequations}\label{eq:A1_5}
\begin{align}
    (1-\kappa^2)\omega_{\lambda}^4+\omega_{\lambda}^2 \bigl[\g^p\gamma_2^p\rho(1-\kappa^2)-1-\rho^2 \bigr] + \rho^2 = 0, \label{eq:A1_5a}  \\
    \g^p = \gamma_2^p\rho \frac{1-\omega_{\lambda}^2(1-\kappa^2)}{\rho^2-\omega_{\lambda}^2(1-\kappa^2)}. \label{eq:A1_5b} 
\end{align}
\end{subequations}
Substituting Eq.~\eqref{eq:A1_5b} into Eq.~\eqref{eq:A1_5a} yields 
\begin{align} \label{eq:A1_6}
    &\omega_{\lambda}^6(1-\kappa^2)^2 - \omega_{\lambda}^4 \big[(1-\kappa^2)(1+2\rho^2)-\rho^2(\gamma_2^p)^2(1-\kappa^2)^2\big] \nonumber \\
    &\quad -\omega_{\lambda}^2 \biggr\{\rho^2(1-\kappa^2) \bigl[(\gamma_2^p)^2-1\bigr] -\rho^2-\rho^4\biggr\} - \rho^4=0.
\end{align}
Eq.~\eqref{eq:A1_6} is the characteristic polynomial whose solutions are the real steady-state modes. Since $R_{eff} = (\g^p)^{-1}\sqrt{L_1/C_1}$, Eq.~\eqref{eq:A1_5b} can be rewritten to define $R_{eff}$ as
\begin{equation}\label{eq:A1_7}
   R_{eff} = \frac{R_2}{\chi^2}\frac{\rho^2-\omega_{\lambda}^2(1-\kappa^2)}{1-\omega_{\lambda}^2(1-\kappa^2)}.
\end{equation}
In the $\mathcal{PT}$-symmetric case, $\rho = \chi = 1$. Assuming $\omega_{\lambda}\ne \pm\sqrt{1/(1-\kappa^2)}$, Eq. \eqref{eq:A1_5b} yields $g_{(l,h)}^p = \gamma_2^p$. Substituting this condition into Eq. \eqref{eq:A1_5a}, we find the real modes $\omega_{(l,h)}$ in Eq.~\eqref{subeq:4c}. The third mode, $\omega_{(m)}$, is found by noting that Eq.~\eqref{eq:A1_6} assumes the denominator of Eq.~\eqref{eq:A1_5b} cannot be zero, which occurs at $\omega_{(m)} = \pm\sqrt{1/(1-\kappa^2)}$. Back-substituting this into Eq.~\eqref{eq:A1_5a} gives the gain, $g_{(m)}^p = (1/\gamma_2^p)\bigl[1/(1-\kappa^2) - 1\bigr]$.

In order to reach steady-state oscillation, the required gain cancels the effective loss seen by the negative resistance. From Fig.~\ref{fig:figappA_ppModes}, we note the presence of mode-splitting above a minimum coupling coefficient, $\kappa_{EP}$; coupling-independent operation is only possible above $\kappa_{EP}$. Based on derivations for $\gamma_2^p$ at the exceptional point in \cite{Chen_Generalized,Schindler_Experimental}, we derive $\kappa_{EP}$ by setting the term under the outer square root in Eq.~\eqref{subeq:4c} greater than or equal to zero, and solving for the conditions that allow this,
\begin{equation}\label{eq:A1_8}
    \kappa_{EP} = \sqrt{1 - \frac{2\bigl[(\gamma_2^p)^2 + 1\bigr] - 2\sqrt{2(\gamma_2^p)^2 + 1}}{(\gamma_2^p)^4}}, \quad\quad \gamma_2^p \ge 0.
\end{equation}
As shown in Fig.~\ref{fig:robustness}(a), $\kappa_{EP}\le\gamma_2$ when $\rho = 1$.  For $\kappa>\kappa_{EP}$, the effective resistance is coupling independent and constant at $R_{eff} = R_2$ for the two modes, $\omega_{(l,h)}$. Conditions where $\kappa \le \kappa_{EP}$ are henceforth defined as under-coupled; in this case, the only real mode is $\omega_{(m)}$ and $g_{(m)}$ (and hence $R_{eff,(m)}$) is coupling-dependent.

\subsection{Middle Mode Energy Conservation}
\label{sec:Energy_Cons}

\begin{figure}[t!]
    \centering
    \includegraphics[scale = 0.6]{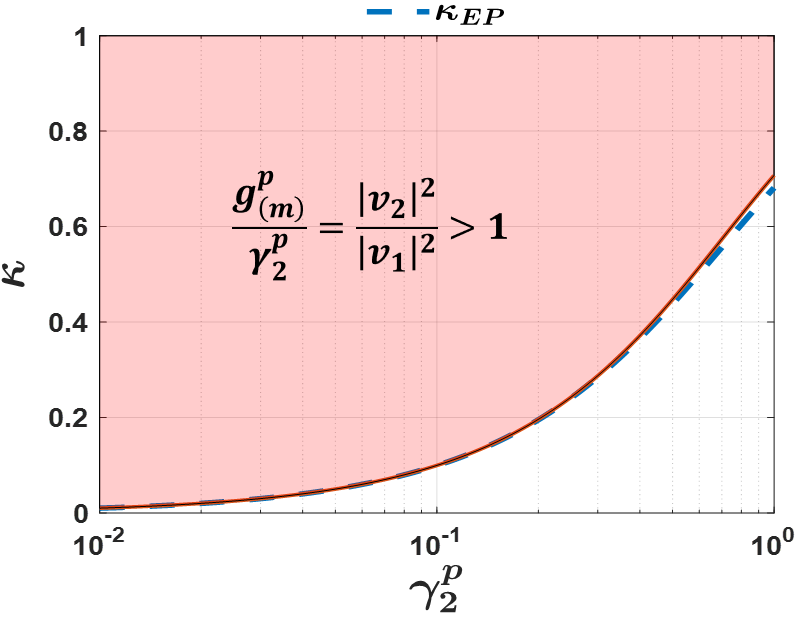}
    \caption{\label{fig:AppA_Energy_Cons} Parametric plot of $\kappa_{EP}$ vs. $\gamma_2^p$. The shaded region shows where $g_{(m)}^p/\gamma_2^p>1$, which is identical to the region of exact $\mathcal{PT}$-symmetry as specified by $\kappa_{EP}$. The middle mode, $\omega_{(m)}$, violates energy conservation in exact $\mathcal{PT}$-symmetry and can therefore only exist in the under-coupled region ($\kappa < \kappa_{EP}$).}
\end{figure}

\noindent The middle mode does not obey energy conservation under mode-splitting and can therefore never sustain oscillations. We begin showing this by assuming sinusoidal steady-state dependence, $v_1\propto e^{i\omega_{\lambda}\tau}$ and $v_2\propto e^{i\omega_{\lambda}\tau}$. In this case, after substituting $q_n=Cv_n$ in Eqs.~\eqref{eq:A1_3}, the tuned ($\rho=\chi=\mu=1$) coupled-rate equations become,
\begin{subequations}\label{eq:A2_1}
\begin{align}
        -\omega_{\lambda}^2v_1 &= -\frac{1}{1-\kappa^2}v_1 + \frac{\kappa}{1-\kappa^2}v_2 + i\omega_{\lambda}\g^pv_1 \label{subeq:A2_1a} \\
        -\omega_{\lambda}^2v_2 &= -\frac{1}{1-\kappa^2}v_2 + \frac{\kappa}{1-\kappa^2}v_1 - i\omega_{\lambda}\gamma_2^pv_2. \label{subeq:A2_1b}
\end{align}
\end{subequations}
We re-arrange Eqs.~\eqref{eq:A2_1} to give two unique expressions for the voltage intensity ratio,
\begin{subequations}
\begin{align}
        \frac{|v_1|^2}{|v_2|^2} &= \frac{\frac{\kappa^2}{(1-\kappa^2)^2}}{\bigl(\frac{1}{1-\kappa^2} - \omega_{\lambda}^2\bigr)^2 + \omega_{\lambda}^2(\g^p)^2} \\
        \frac{|v_1|^2}{|v_2|^2} &= \frac{\bigl(\frac{1}{1-\kappa^2} - \omega_{\lambda}^2\bigr)^2 + \omega_{\lambda}^2(\gamma_2^p)^2}{\frac{\kappa^2}{(1-\kappa^2)^2}}.
\end{align}
\label{eq:A2_2}
\end{subequations}
From Eqs. \eqref{eq:A2_2}, we can solve for the ratio of $\g^p$ and $\gamma_2^p$,
\begin{equation}\label{eq:A2_3}
        \frac{\g^p}{\gamma_2^p} = \frac{|v_2|^2}{|v_1|^2}\sqrt{\frac{\frac{\kappa^2}{(1-\kappa^2)^2} - \frac{|v_1|^2}{|v_2|^2}\biggl[\frac{1}{1-\kappa^2} - \omega_{\lambda}^2\biggr]^2}{\frac{\kappa^2}{(1-\kappa^2)^2} - \frac{|v_2|^2}{|v_1|^2}\biggl[\frac{1}{1-\kappa^2} - \omega_{\lambda}^2\biggr]^2}}.
\end{equation}
Substituting the middle eigenfrequency, $\omega_{(m)} = \sqrt{1/(1-\kappa^2)}$, into Eq. \eqref{eq:A2_3} gives,
\begin{align}\label{eq:A2_4}
        \frac{g_{(m)}^p}{\gamma_2^p} = \frac{|v_2|^2}{|v_1|^2}.
\end{align}
Since $|v_1|^2$ and $|v_2|^2$ correspond to the energy stored in either resonator, conservation of energy stipulates $|v_2|^2 \le |v_1|^2$, or, $g_{(m)}^p/\gamma_2^p \le 1$. Therefore,
\begin{equation}\label{eq:A2_5}
    \frac{|v_2|^2}{|v_1|^2} = \frac{g_{(m)}^p}{\gamma_2^p} = \frac{1}{(\gamma_2^p)^2}\biggl(\frac{1}{1-\kappa^2} - 1\biggr)<1.
\end{equation}
Eq.~\eqref{eq:A2_5} translates to 
\begin{equation}\label{eq:A2_6}
    \kappa < \sqrt{\frac{(\gamma_2^p)^2}{1 + (\gamma_2^p)^2}}.
\end{equation}
Fig.~\ref{fig:AppA_Energy_Cons} depicts $\kappa_{EP}$ as a function of $\gamma_2^p$ (dashed line) and the red shaded region is where Eq.~\eqref{eq:A2_6} (or equivalently, conservation of energy) is violated. The values of $\kappa$ that satisfy Eq.~\eqref{eq:A2_6} are essentially upper-bounded by $\kappa_{EP}$ and, hence, the exact phase of $\mathcal{PT}$-symmetry. Therefore, $\omega_{(m)}$ violates conservation of energy in the exact phase of $\mathcal{PT}$-symmetry, but not in under-coupled conditions, where it is the only stable mode.

\subsection{Series-Series Resonator Topology Derivations}
\label{sec:ModeDer_SS}

\noindent Now, we consider the series-series resonator topology in Fig.~\ref{fig:AppA_Circuits}(b). Applying KVL, we obtain the Liouvillian,
\begin{equation}\label{eq:A3_1}
    \frac{d}{d\tau}\begin{bmatrix}
    q_1 \\ q_2 \\ \Dot{q_1} \\ \Dot{q_2}
    \end{bmatrix} = \begin{bmatrix}
        0 & 0 & 1 & 0 \\
        0 & 0 & 0 & 1 \\
        -\frac{1}{1-\kappa^2} & \frac{\kappa\mu\chi^2}{1-\kappa^2} & \frac{g_1^s}{1-\kappa^2} & \frac{\gamma_2^s\kappa\chi}{1-\kappa^2}\\
        \frac{\mu\kappa}{1-\kappa^2} & -\frac{\mu^2\chi^2}{1-\kappa^2} & -\frac{g_1^s\kappa\mu}{1-\kappa^2} & -\frac{\gamma_2^s\mu\chi}{1-\kappa^2}\\
      \end{bmatrix}\begin{bmatrix}
    q_1 \\ q_2 \\ \Dot{q_1} \\ \Dot{q_2}
    \end{bmatrix} = \mathcal{L}
    \begin{bmatrix}
    q_1 \\ q_2 \\ \Dot{q_1} \\ \Dot{q_2}
    \end{bmatrix}.
\end{equation}
Note that for series resonators, due to duality, $g_1^s = R_1\sqrt{C_1/L_1}$ and $\gamma_2^s = R_2\sqrt{C_2/L_2}$. The real and imaginary parts of the characteristic equation for this Liouvillian matrix are,
\begin{subequations}\label{eq:A3_2}
\begin{align}
        (1-\kappa^2)\omega_{\lambda}^4+\omega_{\lambda}^2(\g^s\gamma_2^s\rho-1-\rho^2) + \rho^2 = 0, \label{eq:A3_2a}  \\
        \g^s = \gamma_2^s\rho \frac{1-\omega_{\lambda}^2}{\rho^2-\omega_{\lambda}^2}.\label{eq:A3_2b} 
\end{align}
\end{subequations}
Eq.~\eqref{eq:A3_2b} can be rewritten to define $R_{eff}$ as
\begin{equation}\label{eq:A3_3}
    R_{eff} = R_2\mu^2\frac{1-\omega_{\lambda}^2}{\rho^2-\omega_{\lambda}^2}.
\end{equation}
As suggested by Eq.~\eqref{eq:A3_3}, if $\rho = 1$ (exact $\mathcal{PT}$-symmetry), $R_{eff}=R_2\mu^2$, which is independent of $\kappa$, and if $\mu = 1$, then $R_{eff}=R_2$. Substituting Eq.~\eqref{eq:A3_2b} into Eq.~\eqref{eq:A3_2a} yields the following mode solutions,
\begin{equation}\label{eq:A3_4}
    \omega_{(h,l)} = \pm\sqrt{\frac{2 - (\gamma_2^s)^2 \pm \sqrt{\bigl[2-(\gamma_2^s)^2\bigr]^2 - 4(1-\kappa^2)}}{2(1-\kappa^2)}},
\end{equation}
\begin{figure}[t!]
    \centering
    \includegraphics[scale = 0.6]{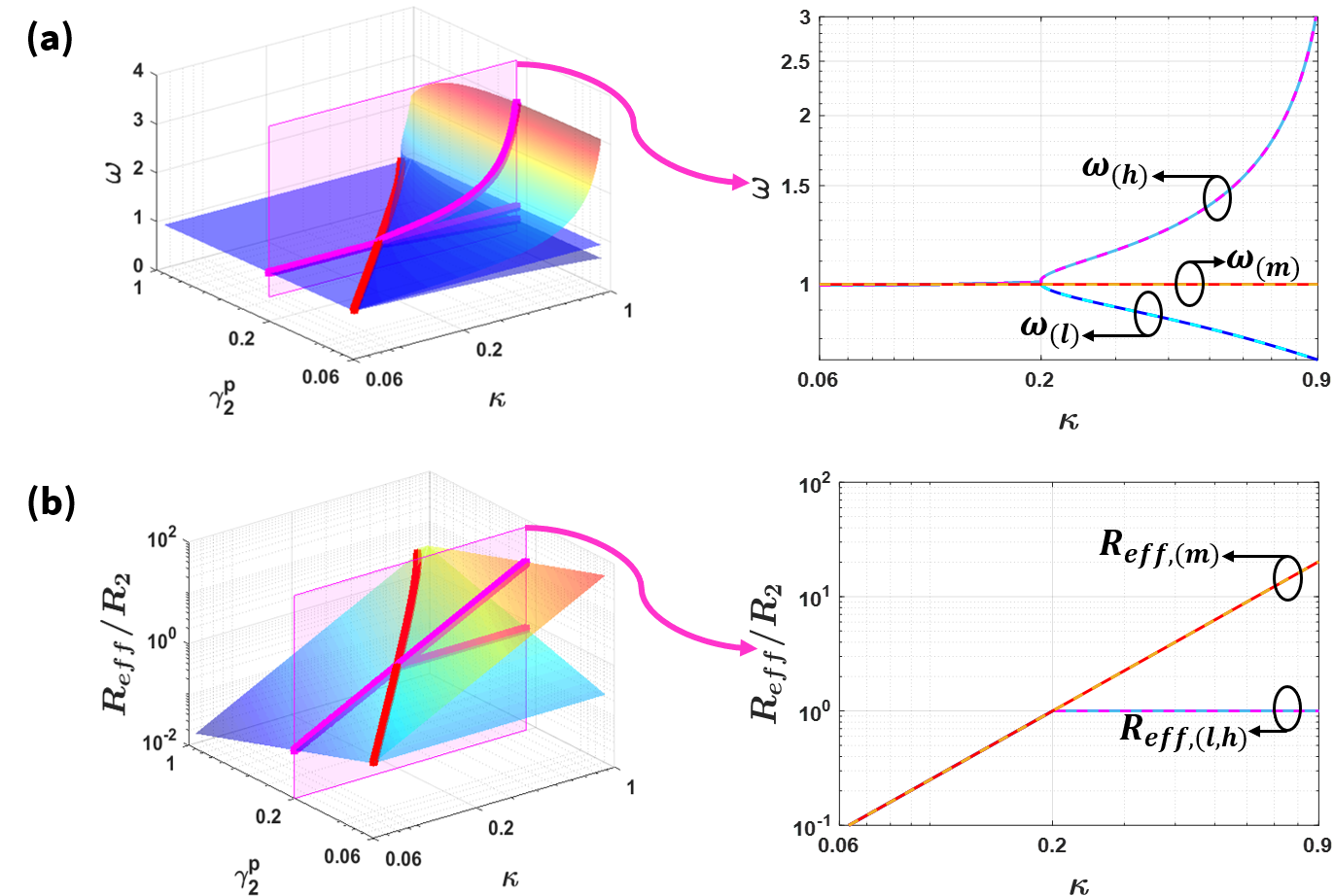}
    \caption{(a) Positive modes and (b) normalized effective resistances corresponding to each mode for the series-series resonator topology in Fig. \ref{fig:AppA_Circuits}(b) and $\rho=1$. Slices show solutions for $\gamma_2^s=0.2$. Mode-splitting past $\kappa_{EP}$ (red markers) allows the lower and upper modes to exhibit a coupling-independent effective resistance. Dashed lines (impedance solutions) confirm Liouvillian solutions (solid lines).  The corresponding circuit parameters are $R_2 = 20.45$~$\Omega$, $L = 2.3$~$\mu$H, and $C = 220$~pF.}
    \label{fig:AppA_ssModes}
\end{figure}

\noindent assuming $\omega_\lambda \ne \pm1$. This $\omega_\lambda$ is the third mode, $\omega_{(m)} =\pm1$; back-substituting this mode into Eq.~\eqref{eq:A3_2a} gives the required gain, $g_{(m)}^s = \kappa^2/\gamma_2^s$.  Fig.~\ref{fig:AppA_ssModes} shows these modes illustrating one real mode for $\kappa<\kappa_{EP}$ and three for $\kappa>\kappa_{EP}$. In the mode-split regime, the third mode, $\omega_{(m)}$, is unstable and will only exist for $\kappa<\kappa_{EP}$. Eq. \eqref{eq:A3_4} gives the following minimum coupling and loss rate range for mode-splitting,
\begin{equation}\label{eq:A3_5}
    \kappa_{EP} = \frac{1}{2}\sqrt{4(\gamma_2^s)^2 - (\gamma_2^s)^4},\quad\quad 0 \le \gamma_2^s \le \sqrt{2}.
\end{equation}
This implies that $\kappa_{EP}\le\gamma_2^s$. Similar to the parallel-parallel case, the effective resistance under mode-splitting is coupling-independent and constant at $R_{eff} = R_2$ for the two modes $\omega_{(l,h)}$ in Eq. \eqref{eq:A3_4} [Fig. \ref{fig:AppA_ssModes}(b)]. 

\subsection{Sensing Range for Parallel-Parallel and Series-Series Resonator Topologies}
\label{sec:Par_Ser}

\begin{figure}[t!]
    \centering
    \includegraphics[scale = 0.46]{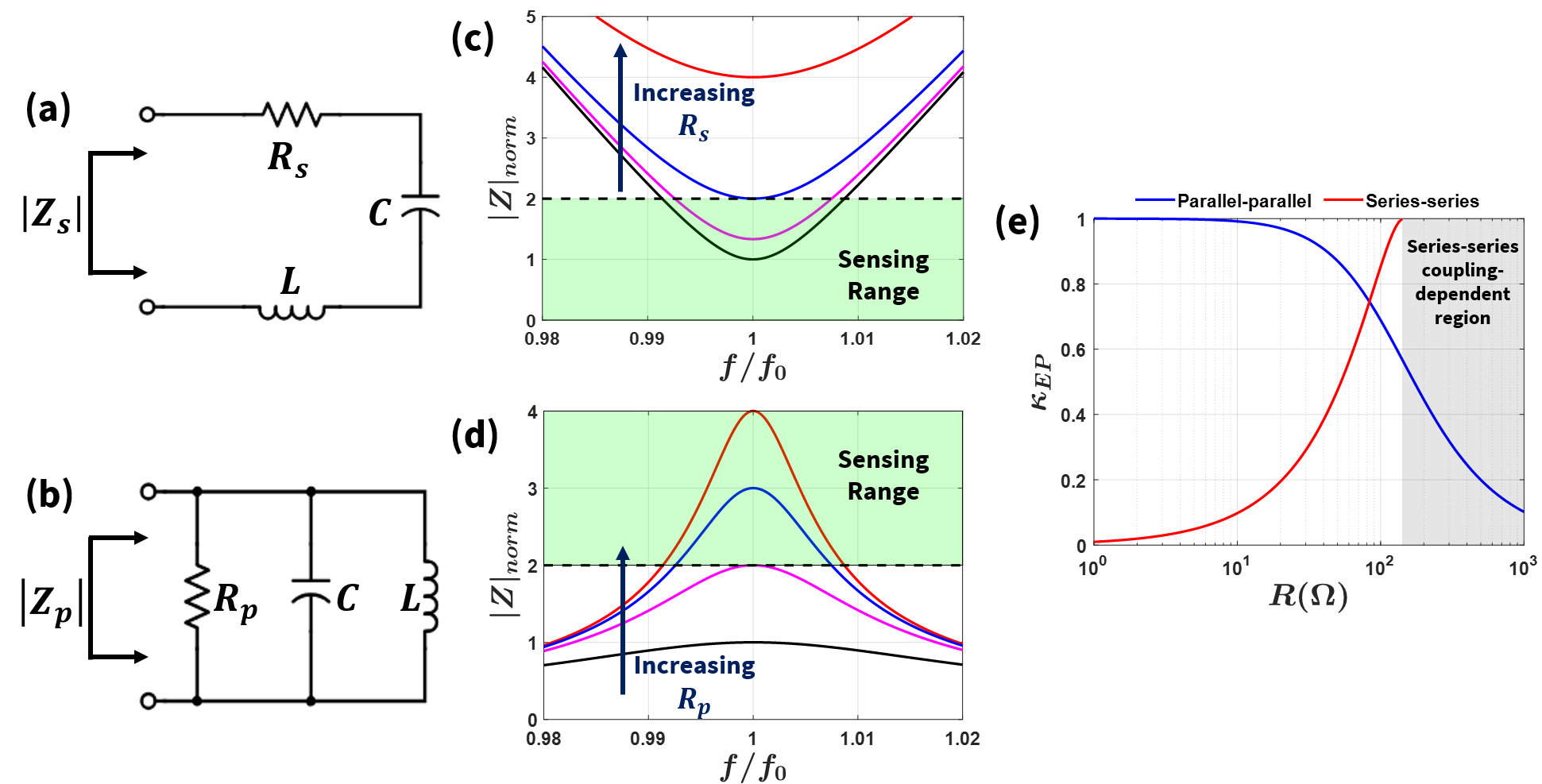}
    \caption{\label{fig:AppA_ParSer} (a) Series and (b) parallel resonators. Resonator impedance vs. frequency and increasing loss parameter for (c) series and (d) parallel resonators. $R_s$ and $R_p$ are calculated using identical resonant frequencies and loss parameters. At resonance, $|Z_s| = R_s$ and $|Z_p| = R_p$. Minimum coupling restricts viable series sensor resistances to the interval, $[0,R_{2,max})$ and viable parallel sensor resistances to the larger, more favorable interval, $(R_{2,min},\infty)$. We assume $R_{2,\cdot} = 2\text{ [a.u.]}$, and normalize the series and parallel impedances to their minimum and maximum values, respectively. (e) Minimum coupling, $\kappa_{EP}$, for the series-series and parallel-parallel resonator topologies assuming the same resistance; note the opposing trend. The circuit parameters are $L=2.3$~$\mu$H and $C=220$pF.}
\end{figure}

\noindent We consider input impedances of the series and the parallel sensor resonators shown in Fig. \ref{fig:AppA_ParSer} (a)-(b). For the series resonator, the complex impedance magnitude, $|Z_s| = \bigl|R_s + i[\omega L - 1/(\omega C)]\bigr|$, is minimized at resonance whereas for the parallel resonator, the complex impedance magnitude, $|Z_p| = \bigl|[1/R_p + i(\omega C - 1/(\omega L)]^{-1}\bigr|$, is maximized at resonance [Fig.~\ref{fig:AppA_ParSer} (c)-(d)]; the inverse relationship is a consequence of duality.

Coupling-independent sensor measurement requires operation beyond $\kappa_{EP}$ which is determined by the loss parameter ($\gamma^p = (R_p)^{-1}\sqrt{L/C}$ and $\gamma^s = R_s\sqrt{C/L}$ for the parallel and series resonators, respectively). This translates to a maximum sensing distance, $d_{max}$, that determines the range of measurable resistance. Since the loss parameters of the two resonators are inversely related, the sensing dynamic range also exhibits opposing trends: $R_{2,max} < R_p < \infty$ and $0 < R_s < R_{2,min}$ for the parallel and series resonators, respectively [Fig. \ref{fig:AppA_ParSer} (c)-(d)]. This suggests that the parallel resonator is well-suited for larger resistance values, while the series resonator is better-suited for smaller resistances. Due to the infinite maximum range, the parallel resonator offers a wider sensing dynamic range than the series resonator. Fig. \ref{fig:AppA_ParSer}(e) further showcases this through the respective $\kappa_{EP}$ for the parallel-parallel and series-series resonator topologies, assuming the same resistance.

\subsection{Series-Parallel and Parallel-Series Resonator Topologies Derivations}
\label{sec:ModeDer_SP_PS}

\begin{figure}[t!]
    \centering
    \includegraphics[scale = 0.6]{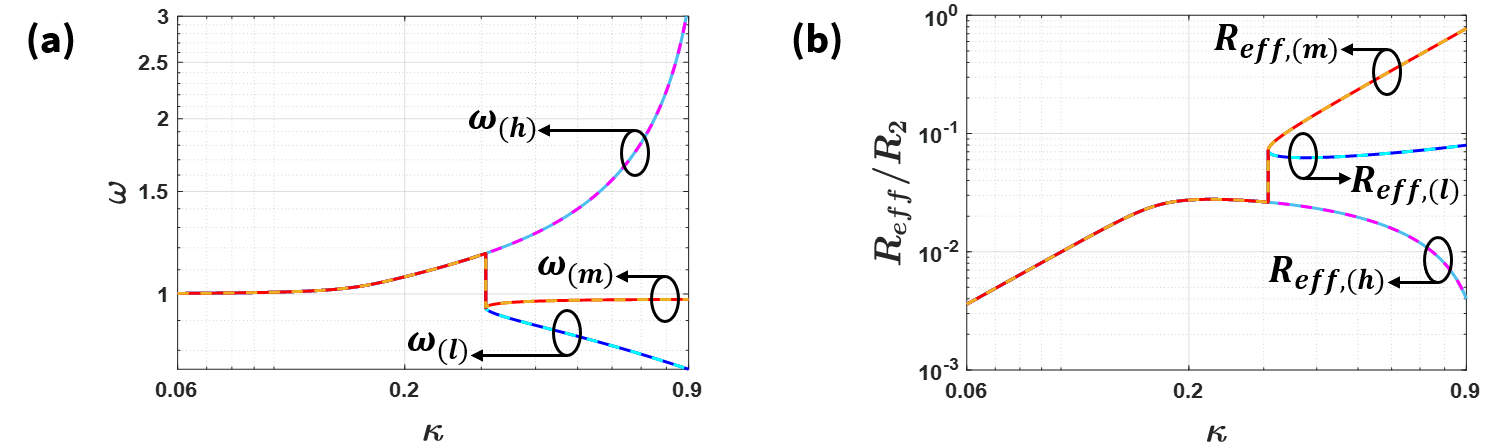}
    \caption{(a) Positive mode and (b) normalized effective resistances corresponding to each mode solution for the series-parallel resonator topology in Fig.~\ref{fig:AppA_Circuits}(c), $\rho=1$, and $\gamma_2^p=0.2$. For $\kappa > \kappa_{EP} = 0.31 >\gamma_2^p=0.2$, mode-splitting is present yet none of the mode solutions exhibits a coupling-independent effective resistance. Dashed lines (impedance solutions) confirm Liouvillian solutions (solid lines). The corresponding circuit parameters are $R_2 = 511.25$~$\Omega$, $L = 2.3$~$\mu$H, and $C = 220$~pF.}
    \label{fig:AppA_spModes}
\end{figure}

\noindent First, we consider the series-parallel topology in Fig. \ref{fig:AppA_Circuits}(c). Using KVL and KCL, we obtain the Liouvillian,
\begin{equation}\label{eq:A5_1}
    \frac{d}{d\tau}\begin{bmatrix}
    q_1 \\ q_2 \\ \Dot{q_1} \\ \Dot{q_2}
    \end{bmatrix} = \begin{bmatrix}
        0 & 0 & 1 & 0 \\
        0 & 0 & 0 & 1 \\
        -\frac{1}{1-\kappa^2} & -\frac{\kappa\mu\chi^2}{1-\kappa^2} & \frac{g_1^s}{1-\kappa^2} & 0\\
        -\frac{\kappa\mu}{1-\kappa^2} & -\frac{\mu^2\chi^2}{1-\kappa^2} & \frac{g_1^s\kappa\mu}{1-\kappa^2} & -\gamma_2^p\mu\chi\\
      \end{bmatrix}\begin{bmatrix}
    q_1 \\ q_2 \\ \Dot{q_1} \\ \Dot{q_2}
    \end{bmatrix} = \mathcal{L}
    \begin{bmatrix}
    q_1 \\ q_2 \\ \Dot{q_1} \\ \Dot{q_2}
    \end{bmatrix}.
\end{equation}
The real and imaginary parts of the characteristic equation for this Liouvillian matrix are,
\begin{subequations}\label{eq:A5_2}
\begin{align}
    (1-\kappa^2)\omega_{\lambda}^4+\omega_{\lambda}^2 (\g^s\gamma_2^p\rho-1-\rho^2) + \rho^2 = 0, \label{eq:A5_2a}  \\
    \g^s = \gamma_2^p\rho \frac{1-\omega_{\lambda}^2(1-\kappa^2)}{\rho^2-\omega_{\lambda}^2}.\label{eq:A5_2b} 
\end{align}
\end{subequations}
Eq.~\eqref{eq:A5_2b} can be re-written to define $R_{eff}$ as,
\begin{equation}\label{eq:A5_3}
    R_{eff} = \frac{L_1}{R_2C_2}\frac{1-\omega_{\lambda}^2(1-\kappa^2)}{\rho^2-\omega_{\lambda}^2}.
\end{equation}
Assuming $\rho=1$ and substituting Eq.~\eqref{eq:A5_2b} into Eq.~\eqref{eq:A5_2a} yields the solutions for real modes under exact $\mathcal{PT}$-symmetry; since $\g^s$ is now a function of $\omega_{\lambda}$, a closed-form solution is no longer instructive. Instead, the numerical solutions are shown in Fig.~\ref{fig:AppA_spModes}.

From Fig.~\ref{fig:AppA_spModes}, we note the presence of mode-splitting only above the minimum coupling coefficient, $\kappa_{EP}$. However, unlike the parallel-parallel and series-series cases where $\kappa_{EP} \le \gamma_2^{p,s}$, minimum coupling occurs for $\kappa_{EP} > \gamma_2^p$ . Additionally, the effective resistance under mode-splitting is no longer constant for any of the three modes. Therefore, for any $\kappa$, the effective resistance is coupling-dependent, rendering the series-parallel resonator topology ineffective for coupling-independent resistive sensing.

Next, we consider the parallel-series resonator topology in Fig.~\ref{fig:AppA_Circuits}(d). Using KVL and KCL, we derive the Liouvillian,
\begin{equation}\label{eq:A6_1}
    \frac{d}{d\tau}\begin{bmatrix}
    q_1 \\ q_2 \\ \Dot{q_1} \\ \Dot{q_2}
    \end{bmatrix} = \begin{bmatrix}
        0 & 0 & 1 & 0 \\
        0 & 0 & 0 & 1 \\
        -\frac{1}{1-\kappa^2} & -\frac{\kappa\mu\chi^2}{1-\kappa^2} & g_1^p & -\frac{\gamma_2^s\kappa\chi}{1-\kappa^2}\\
        -\frac{\kappa\mu}{1-\kappa^2} & -\frac{\mu^2\chi^2}{1-\kappa^2} & 0 & -\frac{\gamma_2^s\mu\chi}{1-\kappa^2}\\
      \end{bmatrix}\begin{bmatrix}
    q_1 \\ q_2 \\ \Dot{q_1} \\ \Dot{q_2}
    \end{bmatrix} = \mathcal{L}
    \begin{bmatrix}
    q_1 \\ q_2 \\ \Dot{q_1} \\ \Dot{q_2}
    \end{bmatrix}.
\end{equation}

\begin{figure}[b!]
    \centering
    \includegraphics[scale = 0.6]{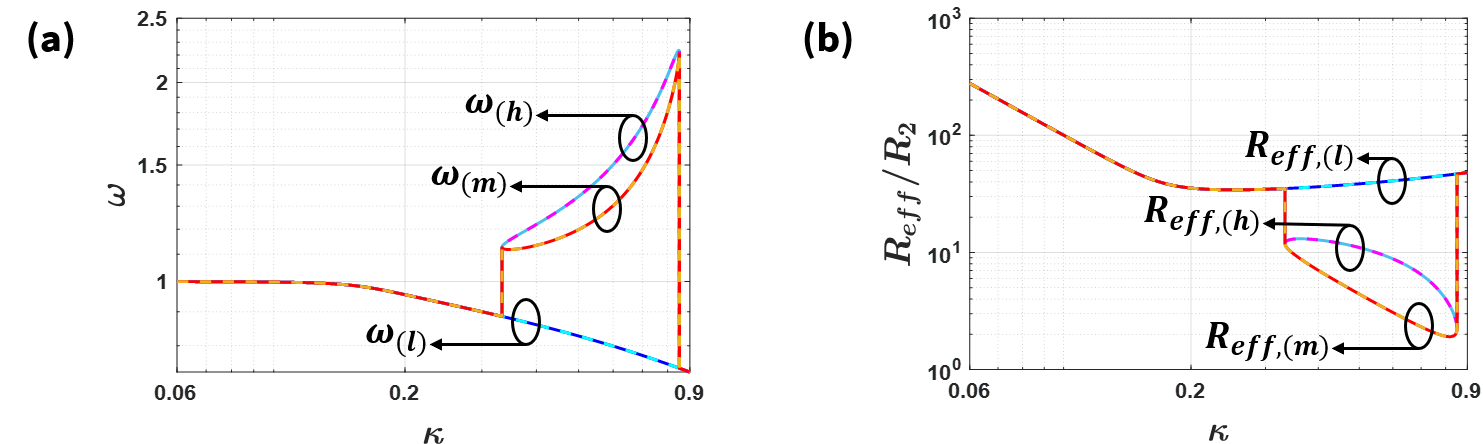}
    \caption{(a) Positive mode and (b) normalized effective resistances corresponding to each mode solution for the parallel-series resonator topology in Fig. \ref{fig:AppA_Circuits}(d), $\rho=1$, and $\gamma_2^p=0.2$. For $\kappa > \kappa_{EP} = 0.33 >\gamma_2^s=0.2$, mode-splitting is present yet none of the mode solutions exhibit a coupling-independent effective resistance. Furthermore, this splitting only happens up to a maximum coupling, $\kappa\approx0.85$. Dashed lines (impedance solutions) confirm Liouvillian solutions (solid lines). The corresponding circuit parameters are $R_2 = 20.45$~$\Omega$, $L = 2.3$~$\mu$H, and $C = 220$~pF.}
    \label{fig:AppA_psModes}
\end{figure}

\noindent The real and imaginary parts of the characteristic equation for this Liouvillian matrix are,
\begin{subequations}\label{eq:A6_2}
\begin{align}
    & (1-\kappa^2)\omega_{\lambda}^4+\omega_{\lambda}^2(\g^p\gamma_2^s\rho-1-\rho^2)+\rho^2 = 0, \label{eq:A6_2a}  \\
    & \g^p = \gamma_2^s\rho \frac{1-\omega_{\lambda}^2}{\rho^2-\omega_{\lambda}^2(1-\kappa^2)}.\label{eq:A6_2b} 
\end{align}
\end{subequations}
Eq.~\eqref{eq:A6_2b} can be rewritten to define $R_{eff}$ as,
\begin{equation}\label{eq:A6_3}
    R_{eff}= \frac{L_2}{R_2C_1}\frac{\rho^2-\omega_{\lambda}^2(1-\kappa^2)}{1-\omega_{\lambda}^2}.
\end{equation}
Assuming $\rho=1$ and substituting Eq.~\eqref{eq:A6_2b} into Eq.~\eqref{eq:A6_2a} yield solutions for the steady-state, real modes under exact $\mathcal{PT}$-symmetry. The $\g^p$ is now a function of $\omega_{\lambda}$; numerical solutions are thus shown in Fig.~\ref{fig:AppA_psModes}.

From Fig.~\ref{fig:AppA_psModes}, we note the presence of mode-splitting only above the minimum coupling coefficient, $\kappa_{EP}$. However, unlike the other topologies, a maximum coupling is observed, past which mode splitting no longer occurs. Additionally, the effective resistance within mode-split region is not constant for any of the three modes. Therefore, for any $\kappa$, the effective resistance is coupling-dependent, rendering the parallel-series resonator topology ineffective for coupling-independent resistive sensing.

\section{Error-Correction Algorithm for Detuned Conditions Using Multiple Discrete Measurements}
\label{sec:MultiMeas_Corr}

\noindent The governing equations for the parallel-parallel resonator topology are given by Eqs.~\eqref{eq:2}. We convert these equations to include the circuit element parameters, arriving at the following functions, $f_1$ and $f_2$,
\begin{subequations}\label{eq:B1}
\begin{align}
    f_1 &= (1-\kappa^2)\omega_{\lambda}^4-\omega_{\lambda}^2\bigg[\omega_1^2+\omega_2^2                     -\frac{\omega_1^2\omega_2^2L_1L_2(1-\kappa^2)}{R_1R_2}\bigg] + \omega_1^2\omega_2^2= 0, \label{eq:B1a}  \\
    f_2 &= R_1 - R_2 \biggl(\frac{\omega_1}{\omega_2}\biggr)^2 \frac{L_1}{L_2} \frac{\omega_2^2-\omega_{\lambda}^2(1-\kappa^2)}{\omega_1^2-\omega_{\lambda}^2(1-\kappa^2)} = 0.\label{eq:B1b} 
\end{align}
\end{subequations}
For each measurement point, assuming non-$\mathcal{PT}$-symmetric conditions, there are three known parameters, $R_1 = R_{eff}$, from the amplitude measurement; $\omega_{\lambda}$, from the frequency measurement; and $\omega_1$, the resonant frequency of the reader, known \emph{a priori} by design. Additionally, we assume identical sensor and reader coils ($L_1 = L_2$). There are, therefore, three unknowns: $\kappa$, $R_2$, and $\omega_2$, temporarily rendering the problem unsolvable. Nonetheless, if two measurements are performed but at two different couplings, then a system of four equations and four unknowns ($\kappa_1$, $\kappa_2$, $R_2$, and $\omega_2$) results and the solution can be found using the generalized Newton-Raphson method for multiple non-linear equations~\cite{eqsolution}. Using Eqs.~\eqref{eq:B1}, we define the functions $f_{11},\, f_{12}$ corresponding to the first measurement and $f_{21},\, f_{22}$ corresponding to the second measurement as,
\begin{align}
       f_{11} &= (1-\kappa_1^2)\omega_{11}^4-\omega_{11}^2\bigg(\omega_1^2+\omega_2^2 -\frac{\omega_1^2\omega_2^2L_1L_2(1-\kappa_1^2)}{R_{11}R_2}\bigg)+\omega_1^2\omega_2^2= 0, \nonumber \\
       f_{12} &= R_{11} - R_2 \biggl(\frac{\omega_1}{\omega_2}\biggr)^2 \frac{L_1}{L_2} \frac{\omega_2^2-\omega_{11}^2(1-\kappa_1^2)}{\omega_1^2-\omega_{11}^2(1-\kappa_1^2)} = 0, \nonumber \\
       f_{21} &= (1-\kappa_2^2)\omega_{21}^4-\omega_{21}^2\bigg(\omega_1^2+\omega_2^2 -\frac{\omega_1^2\omega_2^2L_1L_2(1-\kappa_2^2)}{R_{21}R_2}\bigg)+\omega_1^2\omega_2^2= 0, \nonumber \\
       f_{22} &= R_{21} - R_2 \biggl(\frac{\omega_1}{\omega_2}\biggr)^2 \frac{L_1}{L_2} \frac{\omega_2^2-\omega_{21}^2(1-\kappa_2^2)}{\omega_1^2-\omega_{21}^2(1-\kappa_2^2)} = 0, 
\end{align}
\label{eq:B2}
in which $(R_{11},\, \omega_{11})$ and  $(R_{21},\, \omega_{21})$ correspond to the resistance and the frequency measured from the first and second measurements, respectively. We define $\bm{f} = \begin{bmatrix} f_{11} & f_{12} & f_{21} & f_{22}\end{bmatrix}^T$. This method requires the computation of the Jacobian matrix,
\begin{equation}\label{eq:B3}
    \bm{J} =
  \left[ {\begin{array}{cccc}
   J_{11} & J_{12} & J_{13} & J_{14} \\
    J_{21} & J_{22} & J_{23}&J_{24} \\
     J_{31} & J_{32} & J_{33}& J_{34} \\
     J_{41} & J_{42} & J_{43}& J_{44} \\
  \end{array} } \right] =
  \left[ {\begin{array}{cccc}
   \frac{\partial f_{11}}{\partial R_2} & \frac{\partial f_{11}}{\partial \omega_2} & \frac{\partial f_{11}}{\partial \kappa_1} & 0 \\
    \frac{\partial f_{12}}{\partial R_2} & \frac{\partial f_{12}}{\partial \omega_2} & \frac{\partial f_{12}}{\partial \kappa_1} & 0 \\
     \frac{\partial f_{21}}{\partial R_2} & \frac{\partial f_{21}}{\partial \omega_2} & 0 & \frac{\partial f_{21}}{\partial \kappa_2} \\
     \frac{\partial f_{22}}{\partial R_2} & \frac{\partial f_{22}}{\partial \omega_2} & 0 & \frac{\partial f_{22}}{\partial \kappa_2} \\
  \end{array} } \right].
\end{equation}
The vector of initial conditions, $\bm{x}^{(0)} = \begin{bmatrix} R_2 & \omega_2 & \kappa_1 & \kappa_2 \end{bmatrix}^T$, is calculated by assuming $R_2 = 400$~$\Omega$, $\omega_1=\omega_2$, $\kappa_1 = |(\omega_1/\omega_{11})^2-1|$, and $\kappa_2 = |(\omega_1/\omega_{21})^2-1|$. We can compute
\begin{equation}\label{eq:B4}
    \bm{\Delta x}^{(0)} = -\big[\bm{J}^{(0)}\big]^{-1}\bm{f\big(x^{(0)}\big)}
\end{equation}
and write $\bm{x}^{(1)} = \bm{x}^{(0)} +  \bm{\Delta x}^{(0)}$ which gives the updated vector $\bm{x}$; the iteration is then continued ten times to achieve a sufficiently small variations in $\bm{x}$ \cite{eqsolution}. This process can be applied to more than two measurements, in which the computed value from the previous measurement is used as initial condition for the next measurement.

\section{Nonlinear Gain Theory}
\label{sec:Gain_Nonlin}

\begin{figure}[b!]
    \centering
    \includegraphics[scale = 0.45]{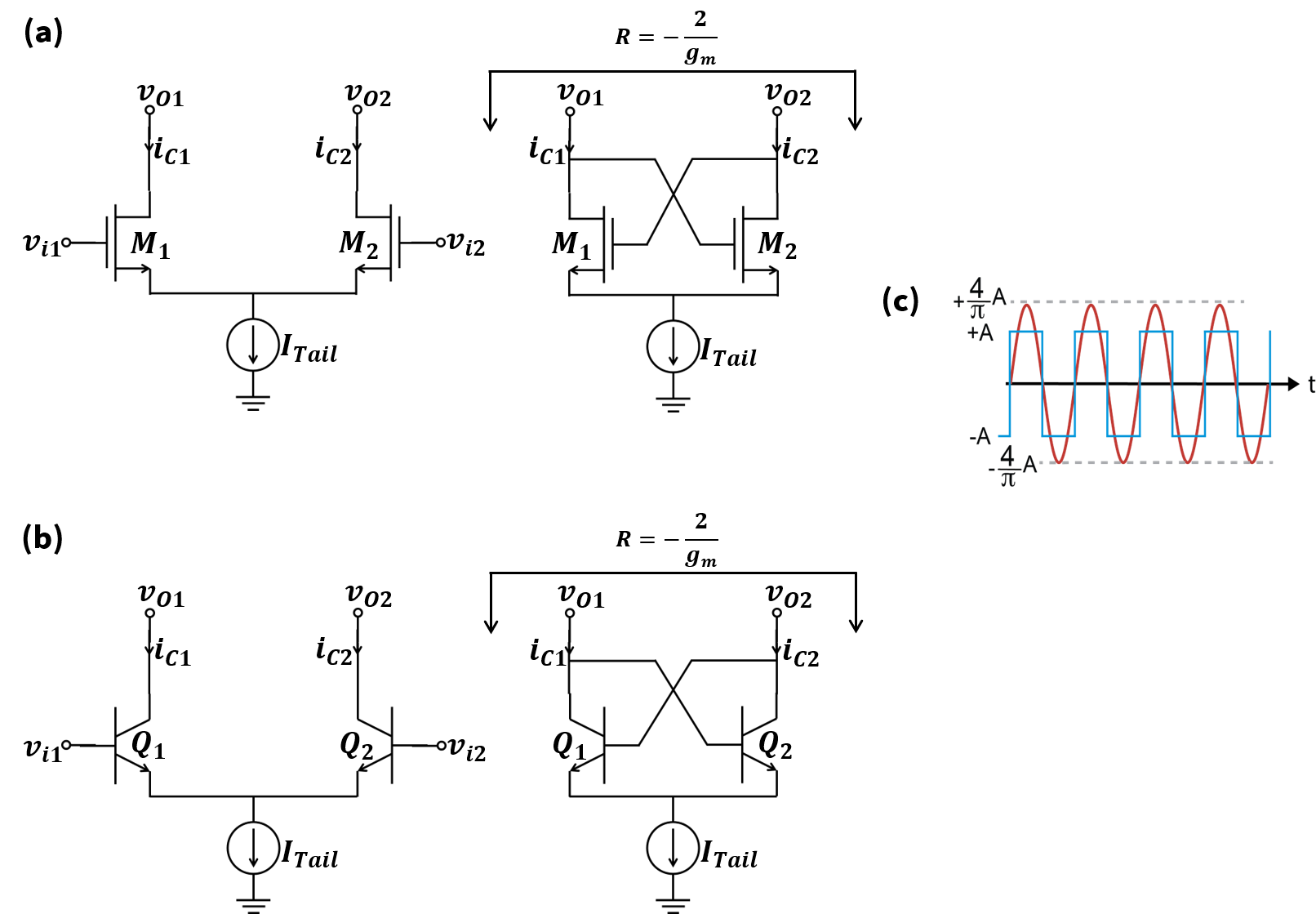}
    \caption{\label{fig:AppC_XCP} (a) MOS and (b) BJT differential-pair and cross-coupled pair implementations. (c) Fourier analysis of the voltage amplitude, $A = R_{eff}I_{Tail}/2$, used in resistive sensing.}
\end{figure}

\begin{figure}[t]
    \centering
    \includegraphics[scale = 0.6]{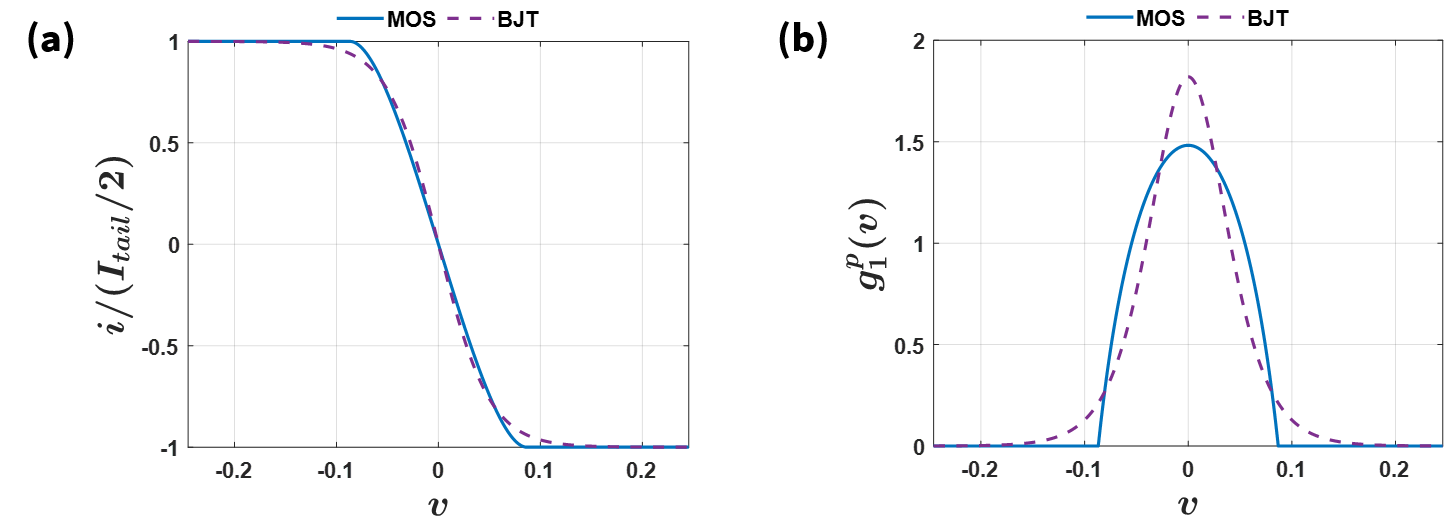}
    \caption{\label{fig:AppC_MOS_BJT_iv} (a) Normalized $i$-$v$ relationships for the MOS and BJT cross-coupled pair implementations. The significant similarity between the two cases allows us to approximate the piece-wise MOS current with the hyperbolic tangent BJT current. (b) Nonlinear gain versus voltage for the  MOS and BJT cross-coupled pair implementations; the non-smooth nature of the MOS gain results in discontinuous second derivatives.}
\end{figure}

\noindent The $g_1(\cdot)$ in Eq.~\eqref{subeq:1a} is described by the nonlinear $i$-$v$ relationship of the negative resistance created by the cross-coupled MOS pair in Fig.~\ref{fig:AppC_XCP}(a). The negative resistance is defined by the differential voltage, $v_{od}=v_{o2}-v_{o1}$, and the current through the drain of $M_1$, $i_{d1}$. We begin by examining this current as a function of the input differential voltage for an MOS differential pair \cite{Gray_Book,Buonomo}. Using KVL and assuming identical transistors with identical threshold voltages, $V_{th1} = V_{th2}$, we write, $v_{i1} - v_{gs1} + v_{gs2} - v_{i2} = 0 \Rightarrow v_{id} = v_{i1} - v_{i2} = V_{od1} - V_{od2}$, where $v_{id}$ is the input differential voltage and $V_{od,n}$ is the overdrive voltage for transistor $M_n$. Here, since the transistors are identical, $\bigl[\mu_n C_{ox}(W/L)\bigr]_1 = \bigl[\mu_n C_{ox}(W/L)\bigr]_2 = 2k$, and we can now define $v_{id}$ as,
\begin{equation}\label{eq:Ca1}
        v_{id} = \sqrt{\frac{i_{d1}}{k}} - \sqrt{\frac{i_{d2}}{k}}.
\end{equation}
We can relate $i_{d1}$, $i_{d2}$, and $I_{Tail}$ using KCL: $i_{d1} + i_{d2} = I_{Tail} \Rightarrow i_{d2} = I_{Tail} - i_{d1}$. Solving for $i_{d1}$ yields,
\begin{equation}\label{eq:Ca2}
        i_{d1} = \frac{I_{Tail}}{2} + v_{id}\sqrt{\frac{kI_{Tail}}{2}}\sqrt{1 - \biggl(\sqrt{\frac{k}{2I_{Tail}}}v_{id}\biggr)^2}.
\end{equation}
The overdrive voltage of each transistor when $v_{id} = -v_{od} = 0$, is $V_{ov} = \sqrt{I_{Tail}/(2k)}$ and the transconductance of the transistors is $g_{m0} = \sqrt{2kI_{Tail}}$. We replace $v_{id} = -v_{od} = -v_1$ and remove the quiescent dc current, $I_{Tail}/2$, to solve for the ac contribution, arriving at
\begin{align}\label{eq:Ca4}
        i_{d1} &= -v_1\frac{g_{m0}}{2}\sqrt{1 - \biggl(\frac{v_1}{2V_{ov}}\biggr)^2} \nonumber \\
        &= v_1\frac{1}{R_{1,0}}\sqrt{1 - \biggl(\frac{v_1}{2V_{ov}}\biggr)^2},
\end{align}
where $R_{1,0} = -2/g_{m0}$ is the initial negative resistance. Eq.~\eqref{eq:Ca4}  assumes both transistors are in saturation \cite{Gray_Book}. However, when $|v_1|>\sqrt{2}V_{ov}$, one transistor enters the cut-off region and the ac current swings at $\pm I_{Tail}/2$ resulting in the following piece-wise $i-v$ relationship,
\begin{equation}\label{eq:Ca5}
        i_{d1} =  \left \{
        \begin{aligned}
            &\frac{I_{Tail}}{2} &, v_1 < -\sqrt{2}V_{ov} \\
            &\frac{v_1}{R_{1,0}}\sqrt{1 - \biggl(\frac{v_1}{2V_{ov}}\biggr)^2} &, -\sqrt{2}V_{ov} \le v_1 \le \sqrt{2}V_{ov} \\
            &-\frac{I_{Tail}}{2} &, \sqrt{2}V_{ov} < v_1.
        \end{aligned}
        \right.
\end{equation}
In the tuned case ($\rho = \chi = \mu = 1$), we know $g_{1,0}^p = -(R_{1,0})^{-1}\sqrt{L/C} = (g_{m0}/2)\sqrt{L/C}$. Therefore, $g_1^p(v_1) =(g_{m}(v_1)/2)\sqrt{L/C}$. Additionally, $g_m(v_1) = \frac{di_{d1}}{dv_1}$~\cite{Gray_Book}, where $v_1=q_1/C_1$. Consequently, $g_1^p(\cdot)$, is the voltage (or, equivalently charge) derivative of the $i$-$v$ relationship,
\begin{equation}\label{eq:Ca6}
        g_1^p(v_1) =  \left \{
        \begin{aligned}
            &0 &, v_1 < -\sqrt{2}V_{ov} \\
            &g_{1,0}^p\frac{1 - \biggl(\frac{v_1}{\sqrt{2}V_{ov}}\biggr)^2}{\sqrt{1 - \biggl(\frac{v_1}{2V_{ov}}\biggr)^2}} &, -\sqrt{2}V_{ov} \le v_1 \le \sqrt{2}V_{ov} \\
            &0 &, \sqrt{2}V_{ov} < v_1.
        \end{aligned}
        \right.
\end{equation}
The piece-wise nature of these $i$-$v$ relationships, however, gives rise to numerical integration issues in MATLAB's ordinary differential equation (ODE) suite when the first or second derivatives are discontinuous. In this case, the second derivative is not continuous at the boundaries $v_1 = \pm\sqrt{2}V_{ov}$. In order to solve this issue, we approximate the discontinuous $i$-$v$ relationship of the MOS cross-coupled pair with that of a BJT pair [Fig. \ref{fig:AppC_XCP}(b)]. To this end, we derive the $i$-$v$ relationship of a BJT cross-coupled pair, in which the collector current is related to $v_1$ through~\cite{Gray_Book,Buonomo},
\begin{equation}\label{eq:Ca7}
    i_{c1} = -\frac{I_{Tail}}{2}\tanh\biggl[\frac{v_1}{2V_T}\biggr],
\end{equation}
where $V_T \approx 25$ mV is the thermal voltage. For the BJT, $g_{m0} = i_{c1}/V_T$ where $i_{c1} = I_{Tail}/2$ is the dc current through transistor $Q_1$. We re-write Eq.~\eqref{eq:Ca7} in terms of the initial design value of the negative resistance, $R_{1,0} = -2/g_{m0}$,
\begin{equation}\label{eq:Ca8}
    i_{c1} = \frac{2V_T}{R_{1,0}}\tanh\biggl[\frac{v_1}{2V_T}\biggr].
\end{equation}
We find the nonlinear gain as the voltage (or, charge) derivative of the $i$-$v$ relationship,

\begin{equation}\label{eq:Ca9}
    g_1^p(v_1) = g_{1,0}^p\sech^2{\biggl[\frac{v_1}{2V_T}\biggr]},
\end{equation}

\noindent where the initial gain may be written in a number of equivalent forms, $g_{1,0}^p = -(R_{1,0})^{-1}\sqrt{L/C} = (g_{m0}/2)\sqrt{L/C} = I_{Tail}/(4V_T)\sqrt{L/C}$. Now, we check how well the MOS $i$-$v$ curve is approximated by the BJT $i$-$v$ curve. From the experimental setup, $k = 0.23611$~A/V$^2$, $I_{Tail} = 1.78$~mA, $L = 2.3$~$\mu$H, $C = 220$~pF, resulting in, $g_{1,0}^p = \sqrt{(kI_{Tail})/2}\sqrt{L/C}\approx 1.5$; using these same numbers with the BJT implementation gives $g_{1,0}^p = I_{Tail}/(4V_T)\sqrt{L/C} \approx 1.8$. The normalized $i$-$v$ and $g_1^p$-$v$ curves in Fig.~\ref{fig:AppC_MOS_BJT_iv} verify that not only do the BJT and MOS cases saturate at the same values of drain/collector current, they also exhibit very similar characteristics in the linear region. Therefore, the hyperbolic tangent response of the BJT implementation is predictive of the MOS circuit behavior; as an added bonus, the function is smooth, avoiding numerical integration issues.

One way to verify the steady-state gain predicted by the fast-time solution [Eq.~\eqref{subeq:2b}], is to examine the steady-state reader-side voltage amplitude. The compressive $i$-$v$ relationship given by Eq.~\eqref{eq:Ca6} results in drain currents that are approximately square waves; the action of the coupled resonators then filters these square waves to their fundamental components~\cite{razavi_book_rfic}. From Fourier Analysis, the resulting sinusoid has an amplitude that is $4/\pi$ larger than the amplitude, $A$, of the square wave [Fig.~\ref{fig:AppC_XCP}(c)]. Assuming the cross-coupled pair sees an effective resistance, $R_{eff}$, presented by the lossy resonator through the coupling mechanism (neglecting parasitic capacitances of the MOS transistors and assuming $\rho = \chi = \mu = 1$), the amplitude of the output differential voltage is,
\begin{equation}\label{eq:Ca10}
    V_1 = \frac{4}{\pi}I_{Tail}\frac{R_{eff}}{2}.
\end{equation}
Here, the factor of 2 is due to the fact that $R_{eff}$ is the differential resistance. The $I_{Tail}$ may be re-written using one of the equivalent forms of $g_{1,0}^p$,
\begin{equation}\label{eq:Ca11}
    V_1 = \frac{8V_Tg_{1,0}^p}{\pi}R_{eff}\sqrt{\frac{C}{L}}.
\end{equation}
In steady state, $R_{eff}=(\g^p)^{-1}\sqrt{L/C}$, resulting in a formula for the expected effective resistance based on the measured voltage amplitude, $V_1$,
\begin{equation}\label{eq:Ca12}
        \frac{\g^p}{\gamma_2^p} = \frac{8V_T}{\pi}\frac{g_{1,0}^p}{V_1\gamma_2^p} \Rightarrow \frac{R_{eff}}{R_2} = \frac{\pi}{8V_T}\frac{V_1\gamma_2^p}{g_{1,0}^p}.
\end{equation}
Fig.~\ref{fig:transient}(a) shows the transient evolution of $R_{eff}$, suggesting that the normalized resistance in Eq.~\eqref{eq:Ca12} settles at unity in steady-state for $\kappa>\kappa_{EP}$.

\section{Circuit implementation and Measurement Setup}
\label{sec:Circ_Imp}

\begin{figure}[b!]
  \centering
    \includegraphics[scale = 0.72]{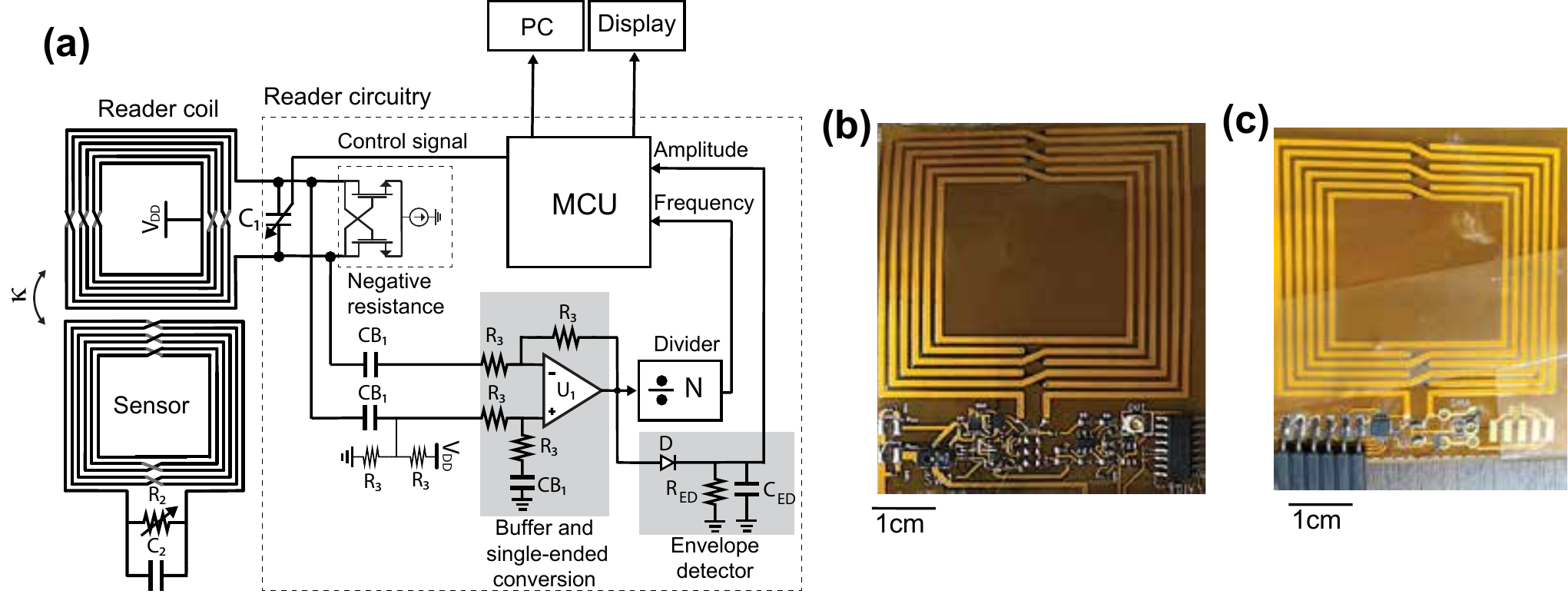}
    \caption{(a) Schematic of the implemented reader circuitry in which the microcontroller unit (MCU) measures the frequency and voltage of self-oscillations. (b)--(c) Reader and sensor implementations using off-the-shelf components on a flexible PCB.}
    \label{fig:AppD_ReaderImp}
\end{figure}

\begin{figure}[t!]
    \centering
    \includegraphics[scale = 0.35]{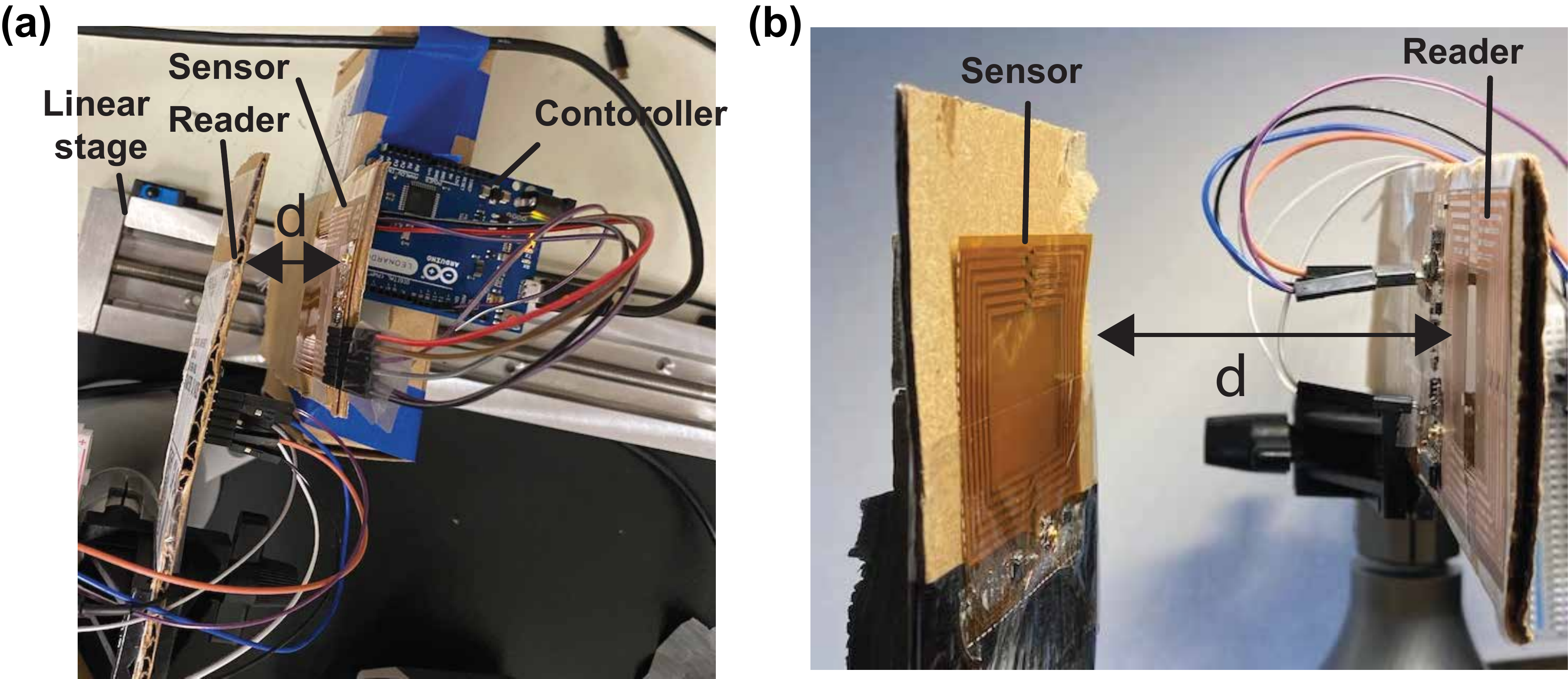}
    \caption{\label{fig:AppD_MeasSetup1} Measurement setups for (a) single-point real-time measurement and (b) multiple-point measurement with imbalanced resonant frequencies.}
\end{figure}

\begin{figure}[h!]
    \centering
    \includegraphics[scale = 0.3]{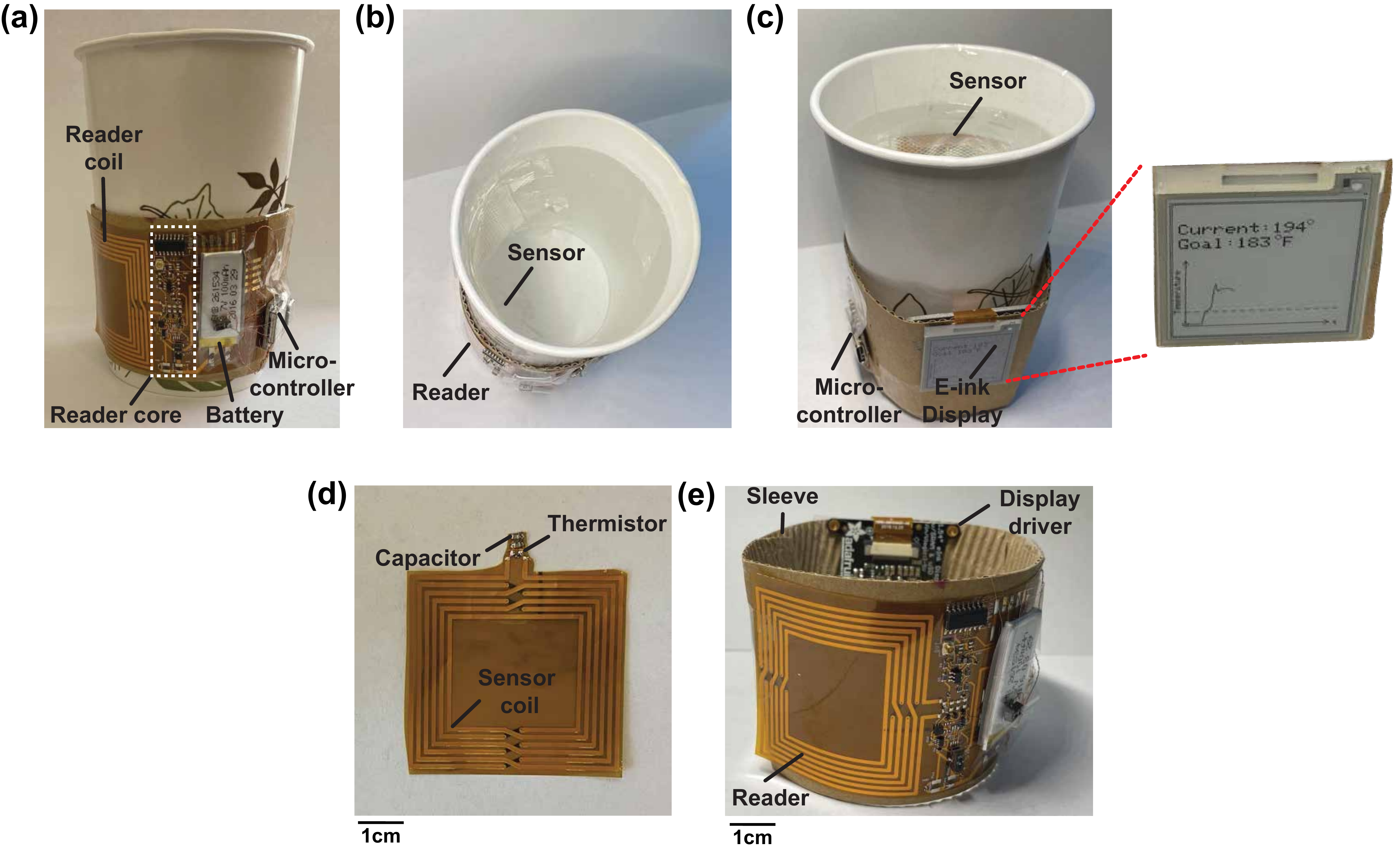}
    \caption{\label{fig:AppD_MeasSetup2} Wireless temperature measurement using a thermistor. (a) The flexible reader mounted on a sleeve around a paper cup, (b) top view of the setup, showing the reader and the sensor inside the cup during wireless sensor measurement, (c) side view of the setup with the E-ink display which shows real-time  temperature profile measured by the reader, (d) fully-passive resistive sensor with the thermistor, and (e) the reader mounted on the sleeve. The sensor and reader are both warped to conform to the shape of the cup.}
\end{figure}

\noindent A prototype is built using off-the-shelf components [Fig.~\ref{fig:AppD_ReaderImp}]; the core reader circuitry consists of the cross-coupled MOS (RUM001L02) pair with a programmable capacitor (NCD2400M) and an inductor ($L = 2.3$~$\mu$ H) implemented using copper traces on a flexible circuit board. The differential oscillation signal is then buffered and converted to single-ended using an op-amp (OPA837) and then applied to a diode-based envelope detector. The frequency is also measured by dividing the signal to within the sampling range of the micro-controller. On the sensor side, the same inductor is used along with a fixed capacitor and a programmable resistor to vary $R_2$. The sensor is mounted on a travel stage and moved horizontally towards the reader from 3~cm to 0.1~cm in 0.1~cm steps while the programmable resistor is varied [Fig.~\ref{fig:AppD_MeasSetup1}(a)]. The measured amplitude and frequency are recorded on a PC. The measurement setup for the error-correction algorithm is shown in Fig.~\ref{fig:AppD_MeasSetup1}(b), in which the distance, d, is varied manually while multiple discrete measurements of $(\omega_\lambda,V_1)$ are made. The measured results are processed in MATLAB. Fig.~\ref{fig:AppD_MeasSetup2} depicts the wireless temperature measurement setup, in which a thermistor (P10574CT-ND) emulates a resistive sensor Fig.~\ref{fig:AppD_MeasSetup2}(c). The sensor resonator is wrapped in an air-tight plastic layer. The reader, shown in Fig.~\ref{fig:AppD_MeasSetup2} operates from a 100~mAh, 3.7~V battery.

\begin{figure}[h!]
    \centering
    \includegraphics[scale = 0.55]{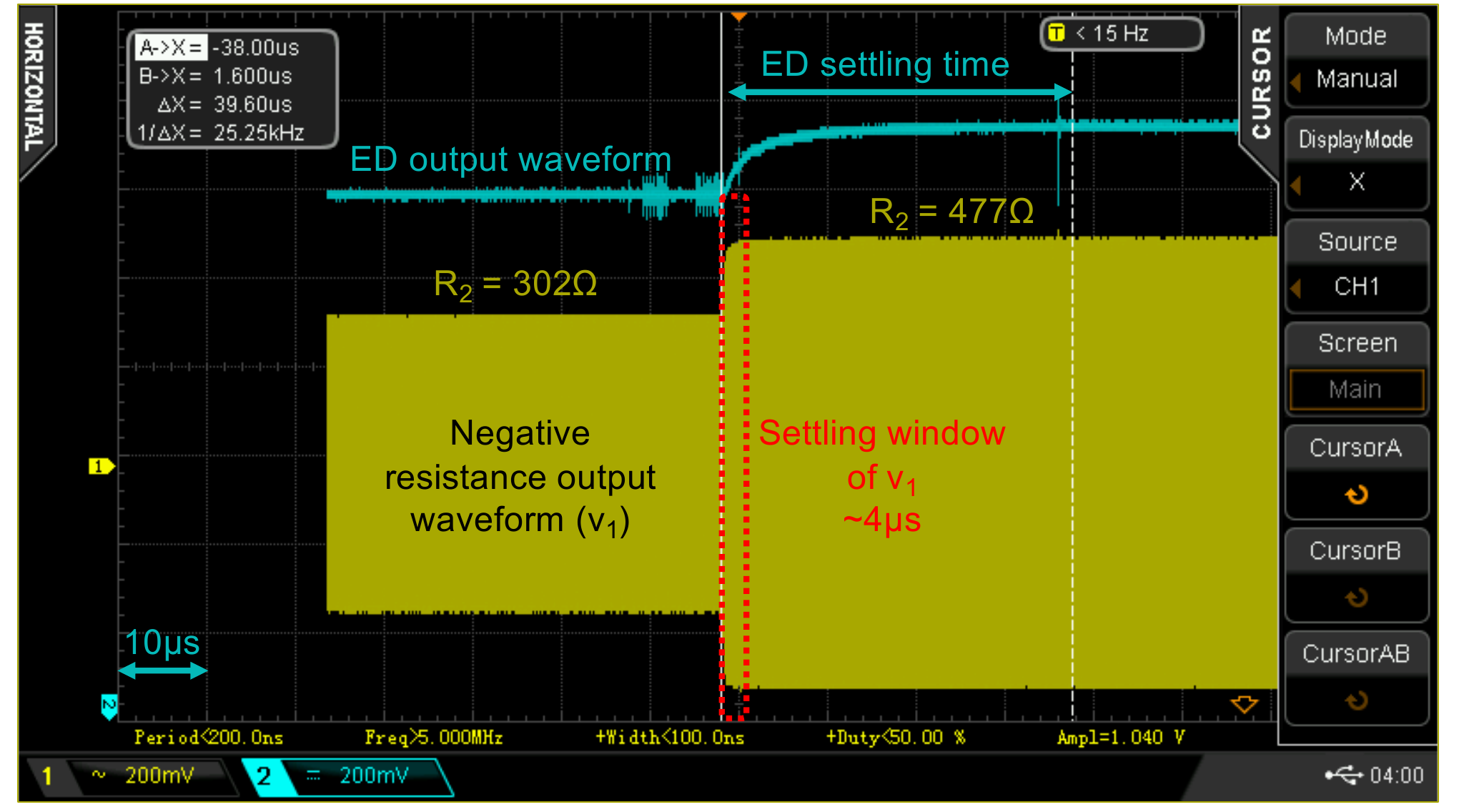}
    \caption{\label{fig:AppD_settling} Settling response of output waveform of the negative resistance (yellow trace) and the output of the ED (blue trace). This suggests that the output of the ED settles within $40\text{ }\mu \text{s}$, while the output waveform of the negative resistance achieves a much faster settling at around $4\text{ }\mu \text{s}$ (around 30 cycles of the reader frequency at $7.1 \text{ MHz}$.)}
\end{figure}

The measured settling behavior of the reader is shown in Fig.~\ref{fig:AppD_settling}, in which the resulting self-oscillation waveform and the measured amplitude are presented when a transition is made from $R_2 = $302~$\Omega$  to $R_2 = $477~$\Omega$. Fig.~\ref{fig:AppD_settling} confirms the real-time sensing capability given the oscillator settles within roughly 4~$\mu$s after the sensor's resistance is altered, matching the fast-settling behavior seen in transient simulations [Fig.~\ref{fig:transient}(b)]. The envelope detector (ED) takes longer to settle (approximately 40~$\mu$s); this is due to the choice of $R_{ED}$ and $C_{ED}$ in Fig.~\ref{fig:AppD_ReaderImp}(a) as a trade-off between settling time and power consumption and attenuation of the higher order harmonics.

%\end{appendix}

%\bibliographystyleNew{plain}
%\bibliographyNew{apssamp}

\end{document}